\documentclass[twocolumn]{aastex7}
\usepackage{amsmath,amssymb,amsfonts}
\usepackage{bm,bbold}
\usepackage{dsfont}
\usepackage{physics}

\newcommand{\mbf}[1]{\mathbf{#1}} 

\begin{document}

\title{Quantum Limits of Exoplanet Detection and Localization}

\author[orcid=0000-0001-8657-3237,gname=Nico,sname=Deshler]{Nico Deshler}
\affiliation{University of Arizona, Wyant College of Optical Sciences, 1630 E University Blvd. Tucson AZ, 85721}
\email[show]{ndeshler@arizona.edu}  

\author[gname=Sebastiaan,sname=Haffert]{Sebastiaan Haffert} 
\affiliation{Leiden Observatory, Leiden University, PO Box 9513, 2300 RA Leiden, The Netherlands6}
\email{strw.leidenuniv.nl}
\author[gname=Amit,sname=Ashok]{Amit Ashok}
\affiliation{University of Arizona, Wyant College of Optical Sciences, 1630 E University Blvd. Tucson AZ, 85721}
\email{ashoka@arizona.edu}

\begin{abstract}
Discovering exoplanets in orbit around distant stars via direct imaging is fundamentally impeded by the combined effect of optical diffraction and photon shot noise under extreme star-planet contrast. Coronagraphs strive to increase the signal-to-noise ratio of exoplanet signatures by optically suppressing light from the host star while preserving light from the exoplanet. However, it is unclear whether direct imaging coronagraphs constitute an optimal strategy for attaining fundamental limits relevant to exoplanet discovery. In this work, we first review the quantum information limits of exoplanet detection and localization characterized by (1) the quantum Chernoff exponent for symmetric hypothesis testing, (2) the quantum relative entropy for asymmetric hypothesis testing, and (3) the quantum Fisher information matrix for multiparameter estimation. We demonstrate that coronagraphs designed to completely suppress light in the fundamental mode of the telescope - while perfectly transmitting higher-order orthogonal modes - indeed achieve these limits in the regime of high star-planet contrasts. Furthermore, we formulate coronagraphs as quantum channels, thus generalizing the classical framework of coronography to the quantum setting. Using this framework, we compare the information-theoretic performance of leading coronagraph designs against the quantum limits. Our analysis indicates that quantum-optimal coronagraphs offer enhanced information efficiency in the sub-diffraction regime compared to leading coronagraph designs and may significantly expand the domain of accessible exoplanets.
\end{abstract}



\section{Introduction}
\label{sec: Introduction}

Exoplanets provide intriguing case-studies for investigating extrasolar dynamics, planet formation, atmospheric science, and potentially the origins of life \cite{Traub:2010_DirectImagingExoplanets,Perryman:2000,Perryman:2003}. Importantly, direct imaging offers the most promising approach for measuring the spectra of an exoplanet which can be used to analyze a planet's atmospheric composition and identify possible bio-signatures \cite{Biller:2013,Currie:2023_DirectImagingSpectroscopy}. However, the search for habitable exoplanets via direct imaging is fundamentally inhibited by the combined effect of optical diffraction and photon shot noise \cite{Biller:2018_ExoplanetAtmosphereImaging, Tsang:2019_ResolvingStarlight}.

For an exoplanet whose angular separation from their parent star resides near or below the Rayleigh resolution limit of the telescope $\sim \lambda/D$, it becomes increasingly difficult to discriminate whether a detected photon originated from the star or from the exoplanet. This uncertainty is a consequence of diffraction through the telescope aperture which destroys the orthogonality between optical fields excited by the star and the exoplanet. The overlap between these optical fields at the image plane, combined with the relative scarcity of exoplanet photons, place fundamental limits on our ability to successfully detect an exoplanet \cite{Helstrom:1969_QuantumDetectionAndEstimationTheory}.

Additionally, fundamental shot noise induced by the Poissonian nature of photon arrivals for thermal sources prohibits us from obtaining exact knowledge of the underlying probability distribution governing our measurement statistics. This means that we cannot simply `subtract out` the contribution from the star in post-processing to reveal the exoplanet. The detriments of this statistical noise are accentuated under high-contrast where shot noise from the star completely overwhelms signal-bearing photons from the exoplanet. Ultimately, our uncertainty in the underlying probability distribution translates into an uncertainty in our estimate of the \textit{parameters} of that distribution, such as the exoplanet's location relative to the star \cite{VanTrees:2004_DetectionAndEstimationTheory}.

Earth-like exoplanets around Sun-like stars are expected to be relatively common, with recent estimates conservatively reporting between 0.37 and 0.6 exoplanets per star using data from Kepler surveys \cite{Bryson:2021_ExoplanetOccurrenceRates}. Most of these exoplanets reside beyond the diffraction limit of existing space-based observatories, requiring indirect detection methods such as radial velocity, astrometry, and/or gravitational microlensing which cannot resolve spectra \cite{Wright:2012_ExoplanetDetectionMethods}.\footnote{The transit method can resolve spectra, however, it requires capturing measurements during a primary or secondary eclipse event where the star, exoplanet, and observatory are collinear \cite{Crossfield:2015_ExoplanetAtmospheres,Madhusudhan:2019_ExoplanetAtmospheres}.} Consider, for example, the James Webb space telescope operating at infrared wavelengths ($D=6.5$ m , $\lambda = 1550$ nm). The statistical population density of giant exoplanets peaks between 1 and 3 astronomical units (AU) away from their parent stars \cite{Fernandes:2019_GiantExoplanetOccurrence}. Meanwhile, Earth-like exoplanets in the habitable zone occupy even tighter orbits at typical distances of 0.1 to 1 AU \cite{Bergsten:2022_KeplerEarthsinHZ}. For solar systems further than approximately $50$ parsecs, all of these orbits fall below the diffraction limit. Moreover, typical star-planet contrast ratios are extremely high, spanning from $10^{5}:1$ for hot giant Jupiter-like planets up to $10^{10}:1$ for colder Earth-like planets \cite{Traub:2010_DirectImagingExoplanets}. Together, the characteristic distance/contrast scales of exoplanets paired with the wave-particle nature of photons conspire to birth a formidable sensing challenge - one that motivates the development of high-resolution telescopes with coronagraphs operating at the fundamental limits imposed by physics.

A coronagraph is a passive linear optical instrument that reduces photon shot noise by rejecting light from the host star prior to photodetection. To this end, various coronagraph designs have been proposed, each with different working principles that lead to associated real-world advantages and disadvantages. Although the landscape of coronagraph designs is vast, all coronagraphs can be modeled as mode-selective optical filters \cite{Guyon:2006,Belikov:2021}, providing a common mathematical framework for comparing their performance. The principal design objective of any coronagraph is to filter out photons residing in the optical modes excited by the star without unnecessarily discarding photons spanning optical modes excited by the exoplanet. Historically, the high-contrast imaging community has quantified the performance of different coronagraphs via operational metrics such as inner working angle and throughput. While these figures of merit have definitive practical utility, to the best of our knowledge, no rigorous information-theoretic analysis of coronagraphs in the context of exoplanet discovery has been undertaken. 

Since the seminal work of Tsang and collaborators \cite{Tsang:2016}, extensive analysis on the quantum information limits of imaging sub-diffraction incoherent source pairs has appeared over the last decade. This body of literature maps directly to the problem of \textit{exoplanet discovery}, which we formally regard as a detection task and a localization task.\footnote{The quantum limits of \textit{exoplanet characterization}, which involve estimating a planet's spectrum \cite{Currie:2023_DirectImagingSpectroscopy,Huang:2023_QuantumLimitsExoplanetSpectroscopy}, mass, radius, orbital dynamics, and density distribution, are equally important threads of research, though they lies beyond the scope of this work.} Indeed, various papers from the quantum information community have already derived information-theoretic bounds for both of these tasks. In this work, our first objective is to consolidate known quantum information bounds on exoplanet discovery and elucidate noteworthy connections between optimal exoplanet detection and localization schemes. Moreover, we hope to offer a self-contained synthesis that bridges insights from the high-contrast imaging communities and quantum information communities. To this end, we examine three quantum information measures: 
\begin{enumerate}
    \item \textit{Quantum Chernoff Exponent (QCE)} - Bounds the asymptotic scaling of the error rate for detecting the presence/absence of an exoplanet planet under symmetric hypothesis testing \cite{Lu:2018,Grace:2022}
    \item \textit{Quantum Relative Entropy (QRE)} - Bounds the asymptotic scaling of false-negative detection rates in under asymmetric hypothesis testing \cite{Huang:2021, Zanforlin:2022}
    \item \textit{Quantum Fisher Information Matrix (QFIM)} - Provides a Cram\'er-Rao bound on the minimum achievable imprecision for an unbiased estimate of the planet position \cite{Rehacek:2017,Prasad:2019_3DSource_APS,Prasad:2020_3DSource_IOPScience,Ang:2017}
\end{enumerate}

Our second objective is to formulate a general quantum model of coronagraphs. Taking inspiration from classical models \cite{Guyon:2006,Belikov:2021}, we define coronagraphs in the quantum setting as pure loss multi-mode bosonic channels followed by a direct imaging measurement. Using this model, we compare the information-theoretic performance of three state-of-the-art instruments: 
\begin{enumerate}
    \item Phase-Induced Amplitude Apodization Complex Mask Coronagraph (PIAACMC)
    \item Vortex Coronagraph (VC)
    \item Perfect Coronagraph (PC)
\end{enumerate}
We find that the PIAACMC and the VC approach the quantum information bounds when star-planet separations are large relative to the diffraction limit, but they diverge from the bounds as the star-planet separation decreases below the diffraction limit. In comparison, we prove that the PC fully saturates the quantum bounds for all star-planet separations at extreme star-planet contrasts. As an alternative to direct imaging coronagraphs, we also consider spatial mode demultiplexing (SPADE) which measures modal projections of the incident optical field. SPADE is also known to saturate all of the quantum information bounds \cite{Huang:2021,Grace:2022,Rehacek:2017}, demonstrating the promise of mode-sorting solutions for future flagship space-telescopes such as NASA's LUVOIR, HabEx, and HWO concepts \cite{Xin:2022,Xin:2023_PhotonicLanternCoronagraphExperiment,Kim:2024_CoherentImaging_w_PL,Desai:2023_PIC_Coronagraphs,Eikenberry:2024_PhotonicImager,Sirbu:2024,Feinberg:2024_HWO}. 

Remarkably, the mathematical expressions for the quantum information bounds on exoplanet detection and localization admit intuitive interpretations, revealing that \textit{exclusive} isolation/rejection of the telescope's fundamental optical mode is critical to realizing an optimal measurement strategy. The quantum bounds also indicate that accessible information persists into deep sub-diffraction star-planet separations, where current observational models predict an abundance of potential exoplanets \cite{Lagrange:2023_Radial,Fernandes:2019_GiantExoplanetOccurrence,Chen:2017,Ning:2018}. We hope that these results inspire the development of realizable quantum-optimal coronagraphs.

\section{Preliminaries}
\label{sec: Preliminaries}
\subsection{Imaging System Model}
\label{sec: Imaging System}
We consider a canonical optical imaging system with a circular pupil of radius $R_0$ and focal length $f$ operating at infinite object conjugate. For a wavelength $\lambda$, the traditional diffraction limit of spatial resolution (Rayleigh Limit) is defined to coincide with the first zero of the Airy pattern located a radial distance $\mathcal{R} =  1.22 \frac{\lambda f}{2 R_0}$ from the optical axis in the focal plane. Let the Cartesian coordinates of the pupil plane and focal plane be $(X',Y')$ and $(X,Y)$ respectively, and define the dimensionless pupil plane $\mbf{u} \equiv \frac{1}{R_0}(X',Y')$ and focal plane $\mbf{r} \equiv \frac{R_0}{\lambda f}(X,Y)$ coordinate vectors. The pupil function $\tilde{\psi}_0(\mbf{u})$ and the point spread function (PSF) $\psi_0(\mbf{r})$ are given by,
\begin{equation}
\tilde{\psi}_0(\mbf{u}) \equiv \frac{1}{\sqrt{\pi}} 
\begin{cases}
    1, & |u| \leq 1\\
    0, & |u| > 1
\end{cases},\qquad 
\psi_0(\mbf{r}) = \frac{J_1(2\pi r)}{\sqrt{\pi} r}
\label{eqn: Pupil Function and PSF}
\end{equation}
where both functions are related by a Fourier transform $\psi_0(\mbf{r}) =\mathcal{F}[\tilde{\psi}_0(\mbf{u})]\equiv \int \tilde{\psi}_0(\mbf{u})e^{-i2\pi \mbf{u} \cdot \mbf{r}} d\mbf{u}$ and satisfy the normalization condition $
\int  |\tilde{\psi}_0(\mbf{u})|^2 d\mbf{u} = \int |\psi_0(\mbf{r})|^2 d\mbf{r} = 1
$. Throughout this work, we refer to the PSF as the `fundamental mode' of the telescope and use $\sigma \equiv \frac{R_0}{\lambda f} \mathcal{R} = 0.61$ as the dimensionless Rayleigh limit.

\subsection{Quantum State Space and Measurements}
\label{sec: State Space and Measurement}
For incoherent quasi-monochromatic thermal sources (see Appendix B of \cite{Tsang:2016}) such as astronomical objects, the quantum state of the optical field is defined on the complex Hilbert space $\mathcal{H} = \mathcal{H}_0 \oplus \mathcal{H}_1$. This space is a direct sum of the multimode vacuum state $\mathcal{H}_0 = \text{span}(\ket{0})$ and single-photon states of the `pixel modes' $\mathcal{H}_1 = \text{span}(\{\ket{\mbf{r}}:\mbf{r}\in\mathbb{R}^2\})$. The state $\ket{\mbf{r}} = \hat{a}^{\dagger}(\mbf{r})\ket{0}$ specifies the creation of a single photon at position $\mbf{r}$ on the image plane, and $\{\hat{a}(\mbf{r}):\mbf{r}\in \mathbb{R}^2\}$ are bosonic annihilation operators satisfying the canonical commutation relation $[\hat{a}(\mbf{r}),\hat{a}^{\dagger}(\mbf{r}')] = \delta(\mbf{r}-\mbf{r}')$. Using this notation, a single-photon state of any square-normalizable transverse spatial mode $\varphi(\mbf{r}):\mathbb{R}^2\rightarrow \mathbb{C}$ can be expressed as,
\begin{equation}
    \ket{\varphi} =  \int \varphi(\mbf{r}) \ket{\mbf{r}}d\mbf{r} \in \mathcal{H}_{1}
\end{equation}
As a result, it can be shown that the inner product between any two single-photon states satisfy,
\begin{equation}
    \braket{\varphi}{\phi} = \int \varphi^{*}(\mbf{r})\phi(\mbf{r})d\mbf{r} \qquad \ket{\phi},\ket{\varphi} \in \mathcal{H}_1
\end{equation}
\noindent Meanwhile the inner product between any single-photon state and the vacuum state is trivially zero $\braket{\phi}{0}=0$. 

For incoherent scenes, it is necessary to work with mixed states $\hat{\rho} \in \mathcal{D}(\mathcal{H})$ where $\mathcal{D}(\mathcal{H})$ is the set of valid density operators on $\mathcal{H}$ satisfying positive semi-definiteness $\hat{\rho}\succeq 0$ and unit-trace $\Tr[\hat{\rho}]=1$. A measurement is defined by a positive operator-valued measure (POVM) $\{\hat{\Pi}_{j}:j \in \mathcal{J} \}$ satisfying,
\begin{equation}
    \hat{\Pi}_{j} \succeq 0 \,\,\, \forall j\in\mathcal{J} \quad \text{and} \quad \sum_{j\in\mathcal{J}} \hat{\Pi}_{j} = \hat{I}
\end{equation}
where $\hat{I}$ is the identity operator on $\mathcal{H}$ and $\mathcal{J}$ represents a generalized (discrete or continuous) index set. In this work, we restrict our attention to von Neumann POVMs comprised of the vacuum projector and projectors on a complete orthonormal basis of single-photon states, 
\begin{equation}
\{ \hat{\Pi}_{\text{vac}} = \dyad{0}\} 
\cup \{ \hat{\Pi}_{j} = \dyad{\pi_j} : j \in  \mathcal{J} \}
\end{equation}
where $\ket{\pi_j} \in \mathcal{H}_1$ and $\braket{\pi_j}{\pi_{j'}}= \delta_{jj'}$. Physically, this class of measurement corresponds to photon counting in a specific spatial mode basis, earning the name Spatial Mode Demultiplexing (SPADE) \cite{Tsang:2016}. Applying such a measurement to any optical field state $\hat{\rho} \in \mathcal{D}(\mathcal{H})$ may yield either a single photon in mode $\pi_j(\mbf{r})$ or no photons at all with probabilities given by the Born rule,
\begin{subequations}
\begin{align}
    &\text{Pr}(\text{1 photon in mode $\pi_j$}) = p_{\pi_j} = \Tr[\hat{\Pi}_{j}\hat{\rho}]\\ 
    &\text{Pr}(\text{no photons}) =  p_{\text{vac}} = \Tr[\hat{\Pi}_{\text{vac}}\hat{\rho}]
\end{align}
\end{subequations}
As a special case, direct imaging corresponds to a POVM involving single-photon projectors $\{ \hat{\Pi}_{\mbf{r}} = \dyad{\mbf{r}} : \mbf{r}\in \mathbb{R}^2\}$.

\subsection{Star-Exoplanet Model}
\label{sec: Star-Exoplanet Model}
Following \cite{Tsang:2016}, we invoke an idealized model where the star and exoplanet are assumed to be quasi-monochromatic point sources, ignoring their finite size and polychromatic emission spectra. We further assume that the scene is effectively static over the measurement period, ignoring the temporal dynamics of the exoplanet around the star. Under these assumptions, the scene is fully characterized by the position of the star $\mbf{r}_s$, the position of the exoplanet $\mbf{r}_e$, and the relative brightness of the exoplanet $b \in (0,1)$. Alternatively, we may parameterize the scene in terms of the center of intensity $\mbf{R} = (1-b)\mbf{r}_{s} + b\mbf{r}_{e}$ and the separation $\mbf{r}_{\Delta} = \mbf{r}_e-\mbf{r}_s$. All subsequent analysis considers the optical axis (origin of the coordinate system) to be aligned with the center of intensity $\mbf{R} = (0,0)$ such that the star and planet positions become,
\begin{equation}
\mbf{r}_s = -b \mbf{r}_{\Delta}, \qquad    
\mbf{r}_e = (1-b)\mbf{r}_{\Delta}
\label{eqn: Star-Planet Positions}
\end{equation}

The parameters $b$ and $\mathbf{R}$ are taken to be known \textit{a priori} such that the localization task amounts to estimating the unknown star-planet separation vector $\mbf{r}_{\Delta}$. Simultaneous estimation of the separation vector, the centroid, and the relative brightness adds substantial complexity to the quantum information analysis which we defer to future studies.\footnote{We point the curious reader to \cite{Rehacek:2017} for an interesting discussion on how the interdependence of these parameters manifests in the QFIM for the one-dimensional counterpart to the exoplanet localization task.}
\begin{figure*}
    \centering
    \includegraphics[width=\linewidth]{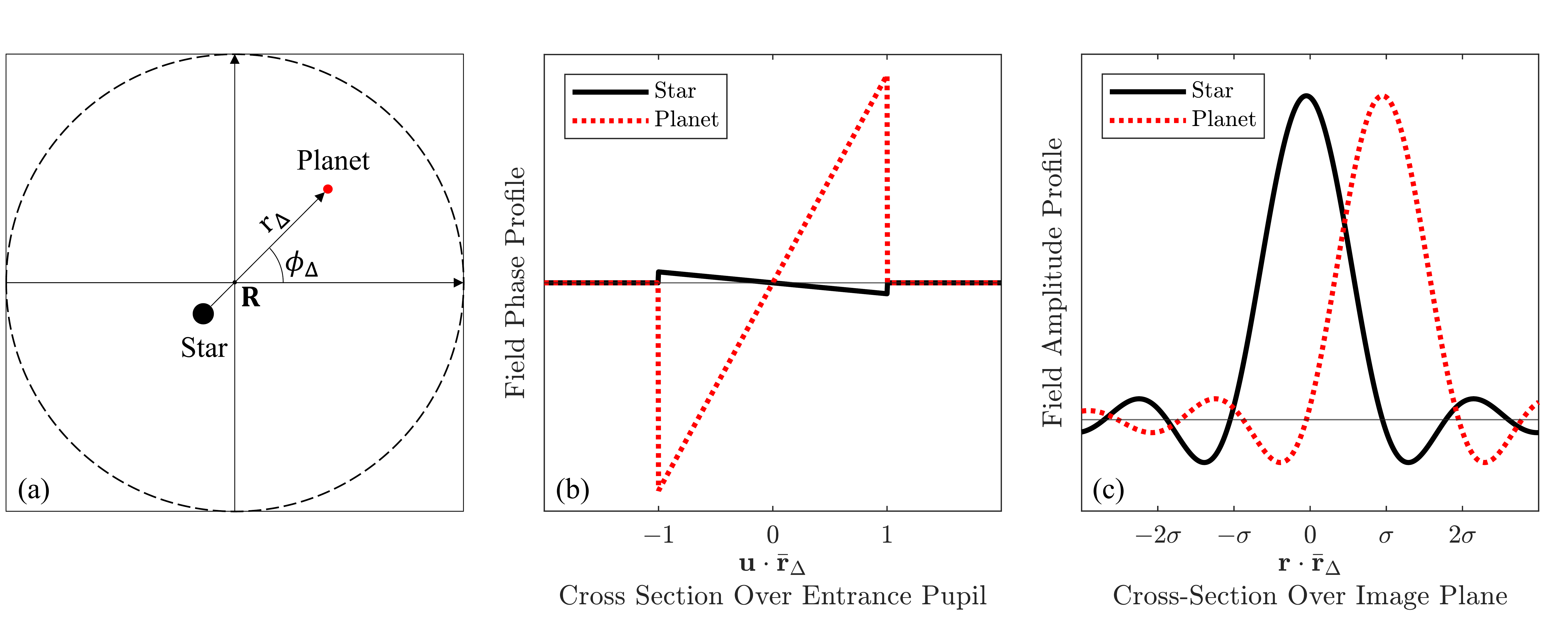}
    \caption{ \textbf{(a)} Geometry of the star-planet model with the optical axis aligned to the center of intensity $\mbf{R}$. \textbf{(b)} Cross-section of the phase profiles over the pupil plane for the field generated by the star and planet independently. \textbf{(c)} Cross-section of the amplitude profiles over the image plane for the fields produced by the star and planet independently. In the limit of high-contrast, the optical axis nears coincidence with the star such that the star purely excites the fundamental mode of the telescope.}
    \label{fig: Star-Exoplanet-System}
\end{figure*}

The optical field produced by the star-planet system over a temporal coherence interval is either in the vacuum state with probability $1-\epsilon$ or occasionally in a single-photon state with probability $\epsilon << 1$. Occupation numbers of more than one photon are assumed to be negligibly rare. The state of the field is thus well approximated by $\hat{\tilde{\rho}} = (1-\epsilon)\dyad{0} + \epsilon \hat{\rho}_1 + \mathcal{O}(\epsilon^2)$ where $\hat{\rho}_1$ is the single-photon mixed state, 
\begin{equation}
    \hat{\rho}_1 \equiv (1-b)\dyad{\psi_0(\mbf{r}_s)} + b\dyad{\psi_0(\mbf{r}_e)}
    \label{eqn: star-exoplanet single-photon density operator}
\end{equation}
encapsulating the possibility that the photon came from either the star or the exoplanet. The states $\ket{\psi_0(\mbf{r}_s)}$ and $\ket{\psi_0(\mbf{r}_e)}$ represent single-photon states of a shifted PSF given by,
\begin{equation}
\ket{\psi_0(\mbf{s})} \equiv \int \psi_0(\mbf{r}-\mbf{s}) \ket{\mbf{r}}d\mbf{r}
\label{eqn: Correlation Function Def}
\end{equation}
Moreover, we define the correlation function between any transverse mode $\varphi(\mbf{r})$ and the shifted PSF state as,
\begin{equation}
    \Gamma_{\varphi}(\mbf{s})\equiv \braket{\varphi}{\psi_{0}(\mbf{s})}
\end{equation}
Conveniently, for any binary pupil (hard aperture) of area $A$, the correlation of an arbitrary mode $\varphi(\mbf{r})$ whose Fourier transform has the same support as the pupil is,
\begin{equation}
    \Gamma_{\varphi}(\mbf{s}) = \frac{1}{\sqrt{A}} \varphi^{*}(\mbf{s}) \, \qquad \forall \varphi : \tilde{\varphi}(\mbf{u})\tilde{\psi}_0(\mbf{u}) = \frac{1}{\sqrt{A}} \tilde{\varphi}(\mbf{u})
\label{eqn: Hard-Aperture Mode Projections}
\end{equation}
a fact which we leverage extensively in the appendix and in numerical simulations.

\section{Quantum Limits of Exoplanet Discovery}

\subsection{Exoplanet Detection}
Exoplanet detection amounts to performing a binary hypothesis test between two possible quantum states of the incident optical field. 
\begin{subequations}
\begin{align}
    \text{Exoplanet Absent } H_0: \hat{\tilde{\rho}}_0 &\approx (1-\epsilon)\dyad{0} + \epsilon \hat{\rho}_{0}\\
    \text{Exoplanet Present }  H_1: \hat{\tilde{\rho}}_1 &\approx (1-\epsilon)\dyad{0} + \epsilon \hat{\rho}_{1}
\end{align}
\end{subequations}
If the exoplanet is absent (null hypothesis $H_0$), then the single-photon component of the field state is given by $\hat{\rho}_0 = \dyad{\psi_0}$ because the optical axis is aligned directly to the star. Otherwise, if the exoplanet is present (alternative hypothesis $H_1$), then the single-photon component is given by $\hat{\rho}_1$ defined in Eq. \ref{eqn: star-exoplanet single-photon density operator}.\footnote{At this point, one may object that the binary hypothesis testing model appears to presume \textit{a priori} knowledge of the exoplanet location $\mbf{r}_{\Delta}$ and relative brightness $b$ since $\hat{\rho}_1$ depends on these parameters. Fortunately, we will show that the quantum-optimal measurement and the maximum likelihood decision criterion are parameter-independent. Thus, we may comfortably proceed with the binary hypothesis testing analysis under the assumption that $\hat{\rho}_1$ is known exactly without affecting the main result.}

Suppose that we are supplied with $M$ copies of one of the two states, either $\hat{\tilde{\rho}}_0^{\otimes M}$ or $\hat{\tilde{\rho}}_1^{\otimes M}$, without being told which. Additionally, let $q_0$ and $q_1 = 1-q_0$ be the prior probabilities on the multi-copy states respectively. In the most general case, the goal of quantum hypothesis testing is to choose a joint measurement (POVM) $\{\hat{\Pi}_{j} : j \in \mathcal{J}\}$ on $\mathcal{H}^{\otimes M}$ and a decision rule $\check{H}:\mathcal{J} \rightarrow \{\check{H}_0,\check{H}_1\}$, which maps a measurement outcome to a guess of the underlying hypothesis, such that the probability of (symmetric or asymmetric) error is minimized. Any pair of POVM and the decision rule can be compressed into a single classification POVM via,
\begin{subequations}
\begin{align}
    \hat{T}_{0} &= \sum_{j\in\mathcal{J}} \hat{\Pi}_{j} \mathbb{1}[\check{H}(j)=\check{H}_0]\\
    \hat{T}_{1} &= \sum_{j\in\mathcal{J}} \hat{\Pi}_{j} \mathbb{1}[\check{H}(j)=\check{H}_1]
\end{align}
\label{eqn: Binary Hypothesis Test POVM}
\end{subequations}
where $\mathbb{1}[\cdot]$ is an indicator function. The total probability of error is thus given by the sum of false-positive (type I) and false-negative (type II) errors weighted by the priors,
\begin{equation}
\begin{split}
P_e &= q_0 P(\check{H_1}|H_0) + q_1 P(\check{H}_0|H_1)\\
    &= q_0 \Tr[\hat{T}_1\hat{\tilde{\rho}}_0^{\otimes M}] + q_1\Tr[\hat{T}_0\hat{\tilde{\rho}}_1^{\otimes M}]
\end{split}
\end{equation}
The Helstrom bound provides a tight lower-bound on the total probability of error achievable with any physically realizable joint measurement,
\begin{equation}
P_{e} \geq P_{e,min} = \frac{1}{2}\bigg[1-||q_1 \hat{\tilde{\rho}}_{1}^{\otimes M} - q_0 \hat{\tilde{\rho}}_{0}^{\otimes M}||\bigg]
\label{eqn: Helstrom Bound}
\end{equation}
where $||\hat{O}|| = \Tr[\sqrt{\hat{O}^{\dagger}\hat{O}}]$ is the trace-norm of a trace-class operator $\hat{O}$. This bound is achieved by POVM that projects onto the positive and negative eigenspectrum of the operator $q_1 \hat{\tilde{\rho}}_1^{\otimes M} - q_{0}\hat{\tilde{\rho}}_0^{\otimes M}$ \cite{Nielsen_Chuang:2010_QuantumInformation}. However, joint measurements act on the full Hilbert space $\mathcal{H}^{\otimes M}$ which corresponds to measuring all received state copies at once (see Appendix \ref{apd: Towards Helstrom Bound}).\footnote{Implementing a joint measurement presumes that there exists a way to perfectly store each copy of the state in a quantum register as they arrive.} In practice, we are often limited to copy-by-copy measurements that act on the smaller Hilbert space $\mathcal{H}$ - such measurements are the focus of subsequent sections. For these measurements, the probability of error decreases exponentially with the number of copies, since, at worst, the total probability of error is given by,
\begin{equation}
P_e = (p_{e})^N = e^{-\xi N}
\end{equation}
where $p_e = e^{-\xi}, \xi>0$ is the single-copy probability of error under a given measurement. A useful metric would be to know the best asymptotic decay rate (largest $\xi$) achieved by optimizing over all possible measurements. 

In the following two subsections, we describe the quantum limits associated with symmetric and asymmetric hypothesis testing reported in \cite{Grace:2022} and \cite{Huang:2021} respectively. We note that the bounds for symmetric and asymmetric hypothesis testing converge in the regime of high contrast. Nevertheless, we include both analyses for completeness.

\subsection{Exoplanet Detection with Symmetric Hypothesis Testing: The Quantum Chernoff Bound}
\label{sec: QCE}

In symmetric hypothesis testing, our goal is to choose a measurement $\{\hat{T}_0,\hat{T}_1\}$ that minimizes the probability of error $P_e$. This problem is called `symmetric' because we look to minimize the combined type-I and type-II error. The optimal probability of error for discriminating whether the field state is $\hat{\tilde{\rho}}_0$ or $\hat{\tilde{\rho}}_1$ asymptotically follows the Quantum Chernoff Bound (QCB) in large $M$ \cite{Audenaert:2007},
\begin{equation}
P_{e,min} \sim e^{-M\tilde{\xi}_Q},\qquad  \tilde{\xi}_{Q} \equiv -\ln(\min_{0\leq t\leq 1} \Tr(\hat{\tilde{\rho}}_{0}^{t}\hat{\tilde{\rho}}_{1}^{1-t}))
\label{eqn: Prob of Error Asympotic}
\end{equation}
where $\tilde{\xi}_{Q}$ is the Quantum Chernoff Exponent (QCE). Under the weak thermal source model, \cite{Grace:2022} showed that,
\begin{equation}
\tilde{\xi}_{Q} \approx \epsilon(1-e^{-\xi_Q}), \qquad \xi_{Q} \equiv -\ln(\min_{0\leq t\leq 1}\Tr(\hat{\rho}_0^{t}\hat{\rho}_1^{1-t}))
\end{equation}
where the $\xi_{Q}$ is the single-photon QCE. Furthermore, when specializing to the sub-diffraction regime $|\mbf{r}_{\Delta}|<<\sigma$ we have (see \cite{Grace:2022} supplementary materials),
\begin{equation}
\tilde{\xi}_{Q} \approx \epsilon \xi_{Q}, 
\end{equation}
Thus, in the context of sub-diffraction exoplanet discovery the QCB depends entirely on $\xi_{Q}$,
\begin{equation}
\xi_{Q} = - \ln \bigg[ (1-b) p_{\psi_0}\big(\mbf{r}_s\big) + b p_{\psi_0}\big(\mbf{r}_e\big)\bigg]
\label{eqn: QCE Single Photon}
\end{equation}
where $p_{\psi_0}(\mbf{s}) = |\Gamma_{\psi_{0}}(\mbf{s})|^2$ is the probability of detecting a photon in the fundamental mode given a point source is located at $\mbf{s}$. In the high-contrast regime where $b\rightarrow 0$, the single-photon QCE tends towards,
\begin{equation}
\boxed{
\xi_{Q}(\mathbf{r}_{\Delta}) \xrightarrow[]{b<<1} b\big[ 1 - p_{\psi_0}(\mbf{r}_{\Delta})\big]
}
\label{eqn: High-Contrast QCE}
\end{equation}
In this limit, the QCE enjoys a simple interpretation - it represents the probability that a photon coming from the exoplanet arrives `outside' of the fundamental mode (i.e., in the orthogonal complement space of the fundamental mode). Thus, the optimal probability of error in the symmetric hypothesis testing decays as,
$$
P_{e,min}\sim e^{-N\xi_{Q}},  \text{ when } b<<1 ,\,\,\, r_{\Delta}<<\sigma
$$
where $N = \epsilon M$ is the total number of photons detected. The high-contrast QCE suggests that separating photons in the fundamental mode from photons in the orthogonal complement space constitutes an optimal strategy for exoplanet detection. Indeed, the so-called Binary SPADE (B-SPADE) measurement,
\begin{equation}
\hat{T}_0 = \dyad{\psi_0},\qquad \hat{T}_1 = \hat{I} -\dyad{\psi_0}
\end{equation}
is quantum-optimal \cite{Grace:2022} and may be applied copy-by-copy. Performing a likelihood ratio test for this measurement amounts to declaring a positive exoplanet detection if any photons are observed in the orthogonal complement space $\hat{T}_1$ within the integration period.

Note that $\hat{T}_0,\hat{T}_1$ are defined independently of the parameters $\mbf{r}_{\Delta}$ and $b$. Therefore, we may apply this POVM to any star-planet configuration. Moreover, under the B-SPADE measurement we have,
\begin{align*}
P(\check{H}_1|H_0) &= \Tr[\hat{T}_1 \hat{\rho}_0]^{N} = 0 \\
P(\check{H}_0|H_1) &= \Tr[\hat{T}_0 \hat{\rho}_1]^{N} = [(1-b)p_0(\mbf{r}_s) + bp_0(\mbf{r}_e)]^{N} 
\end{align*}
We see that the type-I error probability is zero while the type-II error probability shrinks exponentially with the number of photons received. This immediately lends some insight into why the symmetric and asymmetric hypothesis testing cases will be identical; namely because there is no possibility of a false positive.

\begin{table}[!htb]
    \centering
    \resizebox{\columnwidth}{!}{%
    \begin{tabular}{|c|c|}
    \hline
    Telescope Diameter & $6.5\, \text{m}$ \\
    \hline
    Center Wavelength & $1290$ nm  \\
    \hline
    Bandwidth  & $13$ nm ($\sim 1\%$) \\
    \hline
    Visual Magnitude (VM)         & $5.357$ (Prox. Centauri J-Band) \\
    \hline
    Reference Flux (VM=0) & $1589 \times 10^{-26}\, \text{W} \text{m}^{-2} \text{Hz}^{-1}$ \\
    \hline
    Photon Flux & $6 \times 10^7$ Photons $s^{-1}$  \\
    \hline
    \end{tabular}
    }
    \caption{Reference values for calculating exposure times necessary to reach detection and localization performance isocontours in Fig. \ref{fig: QuantumChernoffBound}b and Fig. \ref{fig: QuantumFisherInformation}b respectively.}
    \label{tab: Reference System for Integration Times}
\end{table}

\begin{figure*}
    \centering \includegraphics[width =\linewidth]{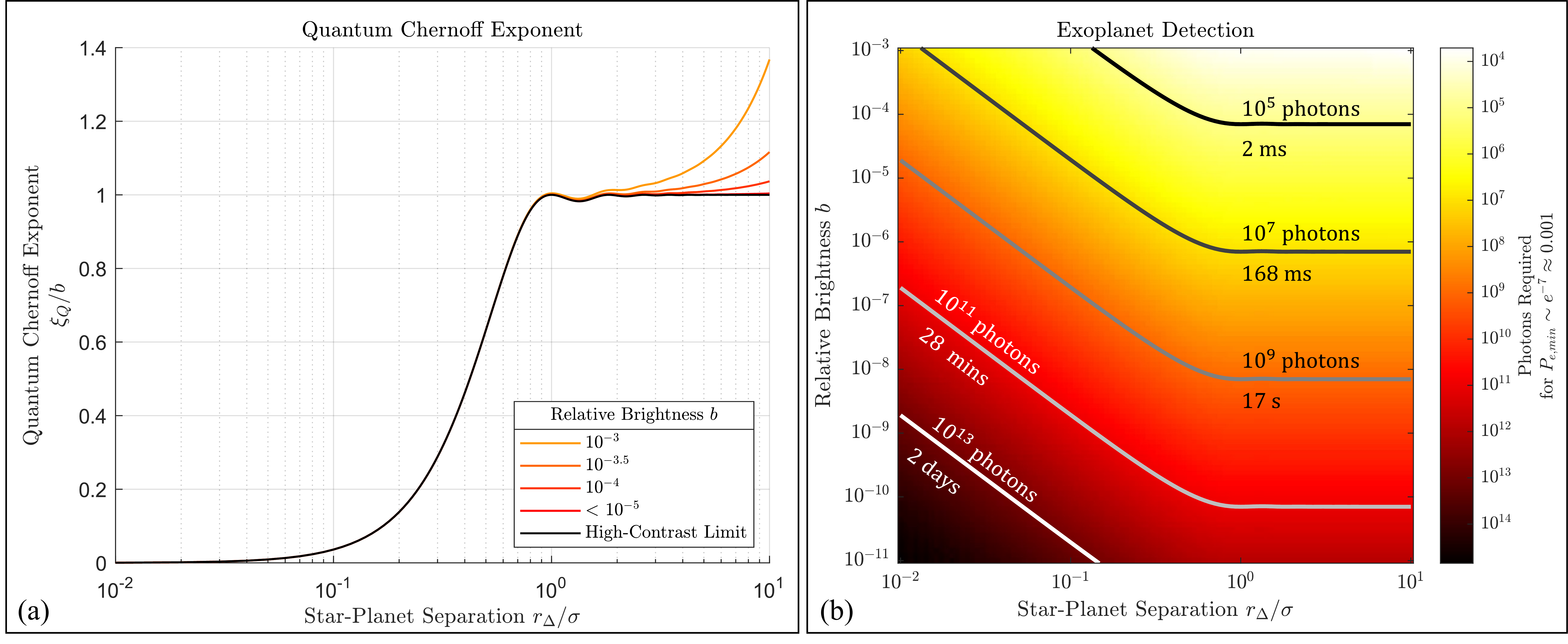}
    \caption{\textbf{(a)} The Quantum Chernoff Exponent (QCE) as a function of the star-planet separation converging towards the high-contrast limit derived in Eq. \ref{eqn: High-Contrast QCE}. \textbf{(b)} A map of the photon requirements for reaching an exoplanet detection confidence of $99.9\%$ assuming quantum-limited performance. The conversion from photon counts to integration times assumes the telescope prescription, stellar magnitude, and waveband defined in Table \ref{tab: Reference System for Integration Times}.}
    \label{fig: QuantumChernoffBound}
\end{figure*}

Fig. \ref{fig: QuantumChernoffBound}(a) shows the single-photon QCE as a function of the star-planet separation converging to the high-contrast limit for a telescope with a circular pupil where $p_{\psi_0}(\mbf{r}_{\Delta}) = (J_{1}(2\pi r_{\Delta})/\pi r_{\Delta})^2$. As the star-planet separation decreases below the diffraction limit $r_{\Delta}<\sigma$, the  single-photon QCE approaches zero. This results from a growing majority of exoplanet photons coupling to the fundamental mode in the regime of small displacements from the optical axis. On the other hand, the  single-photon QCE is maximal wherever the star-planet separation corresponds to a node (zero) of the PSF. At these points, no light from the exoplanet couples into the fundamental mode (a consequence of Eq. \ref{eqn: Hard-Aperture Mode Projections}). Necessarily, the light must couple to higher-order modes if energy is to be conserved. Therefore, exoplanets residing near a node of the PSF may be detected more efficiently. 

The exponential tails of the single-photon QCE observed for moderate contrasts (e.g. $10^{-3}$ to $10^{-4}$) and large star-planet separations $r_{\Delta} > \sigma$ result from starlight coupling to higher-order modes when the system aligned to the center of intensity. Under this alignment, any photon detected outside the PSF mode, whether it be from the star or the exoplanet, is an indication that an exoplanet is present. Fig. \ref{fig: QuantumChernoffBound}(b) quantifies how many photons are required to achieve a designated level of confidence in exoplanet detection for different star-planet contrasts and separation distances using a quantum-optimal instrument.

\begin{figure*}
    \centering
    \includegraphics[width =\linewidth]{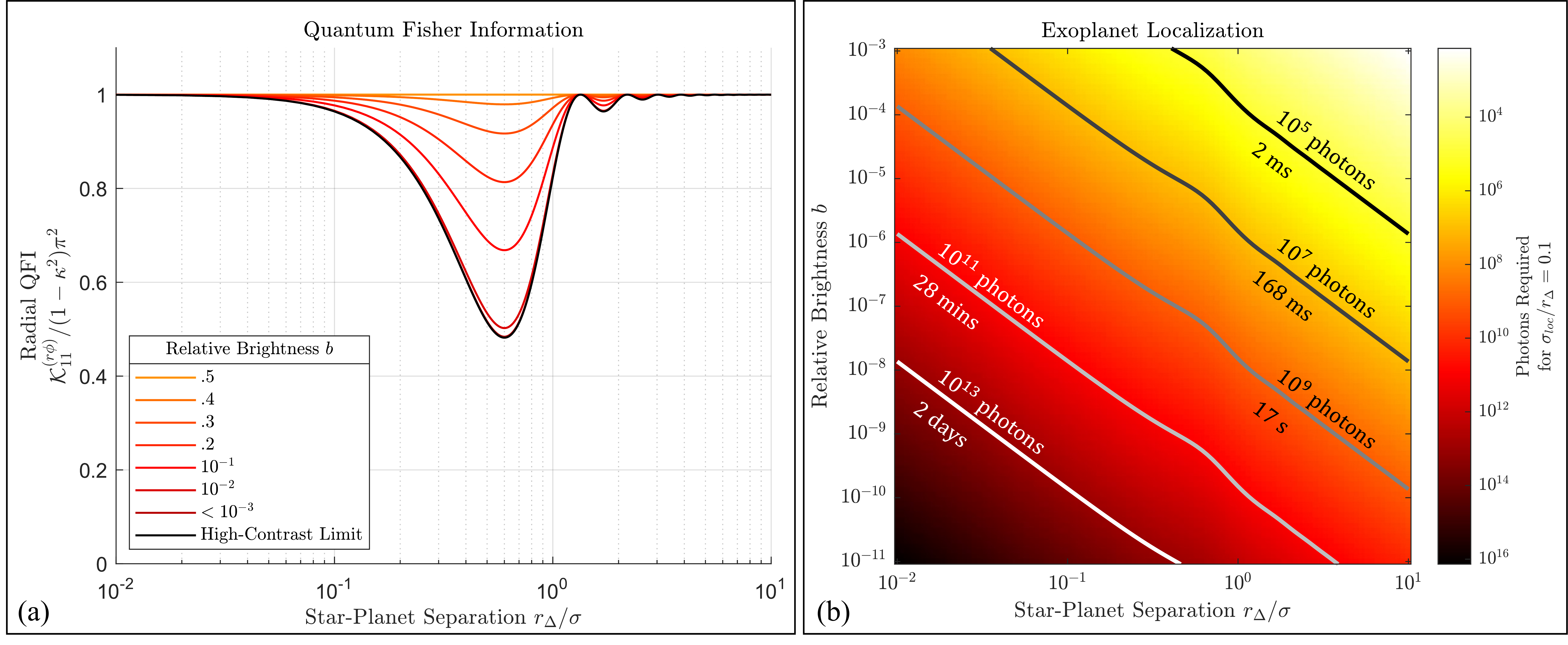}
    \caption{\textbf{(a)} Quantum Fisher Information (QFI) of the radial parameter $r_{\Delta}$ converging to the high-contrast limit expressed in Eq. \ref{eqn: QFIM Polar HC Circular Aperture}. The prominent dip in the QFI appears where the classical Fisher information of the fundamental mode is maximal. The presence of the star suppresses information in the fundamental by a factor of $\sim b$ compared to higher-order modes, resulting in a parameter-dependent reduction of the QFI. \textbf{(b)} Heat map indicating the number of photons required to achieve a relative error in exoplanet localization  of $\sigma_{loc}/r_{\Delta} = 0.1$. The conversion from photon counts to integration times assumes the telescope prescription, stellar magnitude, and waveband defined in Table \ref{tab: Reference System for Integration Times}.}
    \label{fig: QuantumFisherInformation}
\end{figure*}

\subsection{Exoplanet Detection with Asymmetric Hypothesis Testing: The Quantum Stein Lemma}

In asymmetric quantum hypothesis testing, we wish to choose a measurement that minimizes the false-negative rate $P_{e}^{II} \equiv P(\check{H}_0|H_1)$ subject to some required tolerance on the false-positive rate $P_{e}^{I}=P(\check{H}_1|H_0) < \delta$ for $\delta \in (0,1)$. Here we review results from \cite{Huang:2021} where the authors assume that $N$ identical copies of the single-photon states $\hat{\rho}_0$ or $\hat{\rho}_1$ were received. 
Let $P_{e,min}^{II}(\delta)\leq P_{e}^{II}$ be the optimal false-negative error rate optimized over all POVMS on $\mathcal{H}$ that respect $P_{e}^{I}<\delta$. The quantum Stein lemma asserts that the optimal false-negative error scales asymptotically in $N$ as,
\begin{subequations}
\begin{align}
&P_{e,min}^{II}(\delta) \sim e^{-[N \alpha + \sqrt{N\beta}\Phi^{-1}(\delta) + \mathcal{O}(\ln N)]}\\  
&\alpha = D(\hat{\rho}_{0}||\hat{\rho}_{1}) \equiv \langle \ln \hat{\rho}_0 - \ln \hat{\rho}_{1} \rangle_{\hat{\rho}_0}\\
&\beta = V(\hat{\rho}_0 || \hat{\rho}_{1}) \equiv \langle [\ln \hat{\rho}_0 - \ln \hat{\rho}_1 - D(\hat{\rho}_0 || \hat{\rho}_1)]^2 \rangle_{\hat{\rho}_0} 
 \end{align}
 \label{eqn: Quantum Stein Lemma}
\end{subequations}
\noindent where $\langle \hat{A} \rangle_{\hat{\rho}} = \Tr(\hat{A}\hat{\rho})$ is shorthand for the operator expectation, $D(\hat{\rho}_0||\hat{\rho}_1)$ is the Umegaki quantum relative entropy, $V(\hat{\rho}_0||\hat{\rho}_1)$ is the variance of the quantity $(\ln \hat{\rho}_0 - \ln \hat{\rho}_1)$, and $\Phi(x)$ cumulative distribution function for a standard normal random variable \cite{Huang:2021,Wilde:2017_HypothesisTesting}. In appendix \ref{apd: Quantum Stein Lemma}, we provide computable forms for $\alpha$ and $\beta$. 

In the asymptotic limit of many photons $N>>1$ we are concerned only with the dominant term $\alpha$ in the exponential. Previously, \cite{Huang:2021} showed that in the limiting case of high contrast,
\begin{equation}
\boxed{
D(\hat{\rho}_0||\hat{\rho}_1)  \xrightarrow[]{b<<1} b(1-p_{\psi_0}(\mbf{r}_\Delta))
}
\label{eqn: Asymptotic quantum relative entropy}
\end{equation}
which exactly recovers the quantum Chernoff exponent. Thus in the high-contrast regime, $\alpha = \xi_{Q}$ and the quantum Stein lemma converges to the quantum Chernoff bound.

$$
P_{e,min}^{II} \sim P_{e,min} \sim e^{-N\xi_Q}
$$

\subsection{Exoplanet Localization: The Quantum Cramer-Rao Bound}
\label{sec: QFIM}

To localize an exoplanet we consider applying an unbiased estimator over outcomes of measurements on the quantum state. In general, the covariance matrix of an unbiased estimator $\check{\bm{\theta}}$ of the parameters $\bm{\theta} = [\theta_1, \ldots, \theta_{n}]$ characterizing a quantum state $\hat{\rho}(\bm{\theta})$ is lower bounded by the Quantum Cramer-Rao Bound (QCRB)
\begin{equation}
    \text{Cov}[\check{\bm{\theta}}] \geq \bm{\Sigma}_Q = \frac{1}{N} \bm{\mathcal{K}}^{-1}
\label{eqn: def QCRB}
\end{equation}
where $\bm{\mathcal{K}}$ is the Quantum Fisher Information Matrix (QFIM) \cite{Liu:2020}, $N$ is the number of copies of $\hat{\rho}(\bm{\theta})$ measured, and the inequality indicates that $\text{Cov}[\check{\bm{\theta}}] - \bm{\Sigma}_Q \geq 0$ is a positive semi-definite matrix. Thus, given a finite number of copies of the state, any set of parameters (or linear combinations thereof) can be estimated up to a minimum uncertainty determined by the laws of physics. Applying this formalism to the exoplanet localization task, we specialize results from \cite{Prasad:2019_3DSource_APS,Prasad:2020_3DSource_IOPScience} to determine the QFIM for the star-planet separation parameter $\mbf{r}_{\Delta}$. Assuming the optical axis is aligned to the center of intensity, the QFIM for any binary inversion-symmetric pupil $\tilde{\psi}_0(\mbf{u}) = \tilde{\psi}_0(-\mbf{u})$ is given by,
\begin{equation}
    \bm{\mathcal{K}} = \frac{1}{4}(1-\kappa^2)\bigg[\bm{\mathcal{K}}_{1} - \kappa^2 \bm{\mathcal{I}}_{0}\bigg]
\label{eqn: QFI Compact}
\end{equation}
where the constant $\kappa \equiv 1-2b \in [-1,1]$. Additionally, the matrices $\bm{\mathcal{K}}_1, \bm{\mathcal{I}}_0 \in \mathbb{R}^{2\times2}$  are given by,
\begin{subequations}
    \begin{align}
        \mathcal{K}_{1,ij} &= 4 \langle  \partial_{\theta_i} (2\pi \mbf{u}\cdot \mbf{r}_{\Delta} ) \partial_{\theta_j} (2\pi \mbf{u}\cdot \mbf{r}_{\Delta} )\rangle \label{eqn: QFI Single Source}\\ 
        \mathcal{I}_{0,ij} &= 4 \big( \partial_{\theta_i}\Gamma_{\psi_0}(\mbf{r}_{\Delta})\big) \big( \partial_{\theta_j}\Gamma_{\psi_0}(\mbf{r}_{\Delta})\big) \label{eqn: CFI PSF Mode} 
    \end{align}
\end{subequations}
where $(\theta_1,\theta_2)$ are general coordinate components of $\mbf{r}_{\Delta}$ (e.g. Cartesian or Polar) and $\langle \tilde{f}(\mbf{u}) \rangle \equiv \int|\tilde{\psi}_0(\mbf{u})|^2 \tilde{f}(\mbf{u}) d\mbf{u}$ indicates an average over the pupil function. Critically, $\bm{\mathcal{K}}_{1}$ is the constant QFIM associated with localizing a single point source and $\bm{\mathcal{I}}_0$ is the classical Fisher information (CFI) contribution of the fundamental mode.

Note that when the star and planet brightnesses are balanced ($\kappa = 0$), the second term of Eq. \ref{eqn: QFI Compact} vanishes such that $\bm{\mathcal{K}} = \frac{1}{4}{\bm{\mathcal{K}}}_1$, implying that the symmetry of the system gives rise to an equivalence with the single-point source localization problem. The $1/4$ pre-factor arises from the normalization constant of the mixed state describing two balanced incoherent point sources, as opposed to the pure state for single point source. This symmetry recovers the well-known result that the QFI of the separation between two balanced point sources remains constant, irrespective of the separation itself \cite{Ang:2017}. 

In the high-contrast limit, we keep terms of lowest order in $b$ to arrive at the approximation,
\begin{equation}
    \boxed{
    \bm{\mathcal{K}} \xrightarrow[]{b<<1} b \bigg[ \bm{\mathcal{K}}_1 - \bm{\mathcal{I}}_0 \bigg].
    }
\label{eqn: High-Contrast QFIM}
\end{equation}
The QFIM of Eq. \ref{eqn: High-Contrast QFIM} lends itself to an intuitive interpretation - it is the information associated with localizing a single point source if all information in the fundamental mode were unavailable. This interpretation reinforces a long-held intuition within the high-contrast imaging community that the fundamental mode is too noisy to be of value in exoplanet searches since it is dominated by photons from the star. In hindsight, it is remarkable that high-contrast QCE of Eq. \ref{eqn: High-Contrast QCE}, the high-contrast QRE of Eq. \ref{eqn: Asymptotic quantum relative entropy}, and the high-contrast QFIM of Eq. \ref{eqn: High-Contrast QFIM} allude to identical messages despite being information-theoretic measures for different sensing tasks. Namely, that the fundamental mode of the imaging system may readily be dispensed with in the high-contrast regime without incurring information losses about the exoplanet that were otherwise recoverable. This theoretical insight is highly suggestive of optimal coronagraph designs which we explore in the later sections. 

Assuming a circular pupil, the QFIM under polar parameterization of the star-planet separation vector $(\theta_1,\theta_2) =( r_{\Delta},\phi_{\Delta})$ is  explicitly given by,
\begin{equation}
    \bm{\mathcal{K}}^{(r\phi)} = (1-\kappa^2) \pi^2 \Bigg(
    \begin{bmatrix}
    1 & 0 \\
    0 & r^2_{\Delta} 
    \end{bmatrix}
    - \kappa^2 \bigg( \frac{2 \,J_{2}(2\pi r_{\Delta})}{ \pi r_{\Delta}} \bigg)^2 
    \begin{bmatrix}
    1 & 0 \\
    0 & 0 
    \end{bmatrix} 
    \Bigg)
\label{eqn: QFIM polar}
\end{equation}
The QFIM becomes diagonal due to the rotational symmetry of the pupil. When the source brightness are unequal ($\kappa \neq 0$) the  QFI of the radial component $r_\Delta$ acquires a non-trivial parameter-dependence associated with the CFI of the fundamental mode. Meanwhile, the QFI for the angular component $\phi_{\Delta}$ grows quadratically in $r_{\Delta}$, matching a general intuition that angular sensitivity scales with the length of the lever arm. In the high-contrast limit the QFIM for a circular pupil approaches,

\begin{equation}
    \bm{\mathcal{K}}^{(r\phi)} \xrightarrow[]{b<<1} 4\pi^2\, b \Bigg(
    \begin{bmatrix}
    1 & 0 \\
    0 & r^2_{\Delta} 
    \end{bmatrix}
    - \bigg( \frac{2 \, J_{2}(2\pi r_{\Delta})}{ \pi r_{\Delta}} \bigg)^2 
    \begin{bmatrix}
    1 & 0 \\
    0 & 0 
    \end{bmatrix} 
    \Bigg)
\label{eqn: QFIM Polar HC Circular Aperture}
\end{equation}

Fig. \ref{fig: QuantumFisherInformation}(a) shows the convergence of the radial parameter QFI towards the high-contrast limit of equation \ref{eqn: QFIM Polar HC Circular Aperture}. The dominant dip in the radial QFI coincides with the classical information lost to the fundamental mode. To convert the QFIM to a practical measure of the localization precision we consider summing the QCRB variances of the estimation parameters. In polar coordinates, this imprecision is characterized by an uncertainty area patch,

\begin{equation}
    \sigma^2_{loc} \equiv \sigma_{r_{\Delta}}^2 + (r_{\Delta}\sigma_{\phi_{\Delta}})^2 = \frac{1}{N}\bigg[\frac{1}{\mathcal{K}^{(r\phi)}_{11}} + \frac{r_{\Delta}^2}{\mathcal{K}^{(r\phi)}_{22}}\bigg]
    \label{eqn: sigma_loc}
\end{equation}
A related figure of merit is the number of photons needed to localize the exoplanet up to a fraction of the star-planet separation $\frac{\sigma_{loc}}{r_{\Delta}}$ (i.e. relative error). Fig. \ref{fig: QuantumFisherInformation}(b) shows a heat map indicating the number of photons required to reach 10\% relative localization error over a range of star-planet separations and contrasts.

\section{An Information Theoretic Approach to Coronagraphy}

\subsection{Coronagraphs as Quantum Channels}
\label{sec: Coronagraph Quantum Channel}

In the classical setting, a coronagraph is modeled as a passive linear map $C: \mathbb{L}^2\rightarrow \mathbb{L}^2$ between the space of square-normalizable 2D complex functions. In discrete numerical calculations, this map is typically represented as a matrix $\mbf{C} = \mbf{U}_n \mbf{D}_n \cdots \mbf{U}_1 \mbf{D}_1 \mbf{U}_0$, where alternating $\mbf{D}_{i\geq 1}$'s (e.g., diagonal matrices) and $\mbf{U}_{i\geq 1}$'s (e.g. unitary matrices) describe cascaded phase-amplitude modulations of the field inter-spaced with plane-to-plane free-space propagation. The singular-value decomposition (SVD) of the coronagraph map $C= U T V^{\dagger}$ lends the interpretation that a coronagraph is a mode-selective optical filter between a bijective input and output mode space \cite{Belikov:2021, Fontaine:2019_LGMPLC}. The orthonormal input modes $\{\nu_k(\mbf{r}): k\in \mathbb{N}\}$ and orthonormal output modes $\{\mu_{k}(\mbf{r}): k\in \mathbb{N}\}$ are the eigenvectors of unitary operators $V$ and $U^\dagger$ respectively. The singular values $\{\sqrt{\eta_k}\in[0,1]: k\in \mathbb{N}\}$ of $T$ capture the transmissivity between corresponding input and output modes. Thus, any coronagraph is fully characterized by the quantities $\{\eta_k, \nu_k(\mbf{r}),\mu_k(\mbf{r}):k=0,1,\ldots,\infty\}$. Acting the classical coronagraph operator on an arbitrary coherent field $\varepsilon(\mbf{r})$ gives,
\begin{equation}
\varepsilon'(\mbf{r}) = C(\varepsilon(\mbf{r})) = \sum_{k=0}^{\infty} \sqrt{\eta_k}\braket{\nu_k}{\varepsilon}\mu_k(\mbf{r})
\end{equation}

We extend this operation to the quantum setting by defining a coronagraph as a lossy quantum channel given by the completely positive trace preserving map $\mathcal{C}:\mathcal{D}(\mathcal{H})\rightarrow\mathcal{D}(\mathcal{H})$. Sending a state $\hat{\rho}$ through the channel results in the state $\hat{\rho}'$ given by,
\begin{equation}
\hat{\rho}' = \mathcal{C}(\hat{\rho}) = \hat{C}_{1}\hat{\rho}\hat{C}_{1}^{\dagger} + \sum_{k=0}^{\infty} \hat{C}_{0k} \hat{\rho} \hat{C}_{0k}^\dagger + \mel{0}{\hat{\rho}}{0}\dyad{0} 
\label{eqn: Channel Map}
\end{equation}
where $\{\hat{C}_1, \hat{C}_{0k}, \dyad{0}\}$ are the Kraus operators \cite{Ivan:2011,Choi:1975} of the channel,
\begin{equation}
\boxed{
    \hat{C}_{1} = \sum_{k=0}^{\infty} \sqrt{\eta_k} \dyad{\mu_k}{\nu_k} \qquad \hat{C}_{0k} = \sqrt{1-\eta_k} \dyad{0}{\nu_k}
    }
\label{eqn: Kraus Operators}
\end{equation}
Physically, the operator $\hat{C}_{1}$ describes the coherent transmission of modes $\nu_k$ to modes $\mu_k$. The operator $\hat{C}_{0,k}$ describes loss in mode $\nu_k$, resulting in the vacuum state. Critically, for any coronagraph which achieves complete nulling of an on-axis source, the fundamental mode of the telescope necessarily maps to the vacuum state after propagating through channel $\mathcal{C}(\dyad{\psi_0}) = \dyad{0}$ such that, without loss of generality, $\ket{\nu_0} = \ket{\psi_0}$ with $\eta_0 =0$. Applying a direct-imaging measurement to the output state $\mathcal{C}(\hat{\rho})$ gives,
\begin{subequations}
\begin{align}
p_{\mbf{r}} &= \Tr[\hat{\Pi}_{\mbf{r}}\hat{\rho}'] = \sum_{k,k' =0}^{\infty} \sqrt{\eta_k \eta_{k'}} \mel{\nu_k}{\hat{\rho}}{\nu_{k'}} \mu_k(\mbf{r})\mu_{k'}^{*}(\mbf{r})\\
p_{\text{vac}} &= \Tr[\hat{\Pi}_{\text{vac}}\hat{\rho}'] = \mel{0}{\hat{\rho}}{0} + \sum_{k=0}^{\infty}(1-\eta_k)\mel{\nu_k}{\hat{\rho}}{\nu_k}
\end{align}
\end{subequations}

\begin{figure*}
    \centering
    \includegraphics[width=\linewidth]{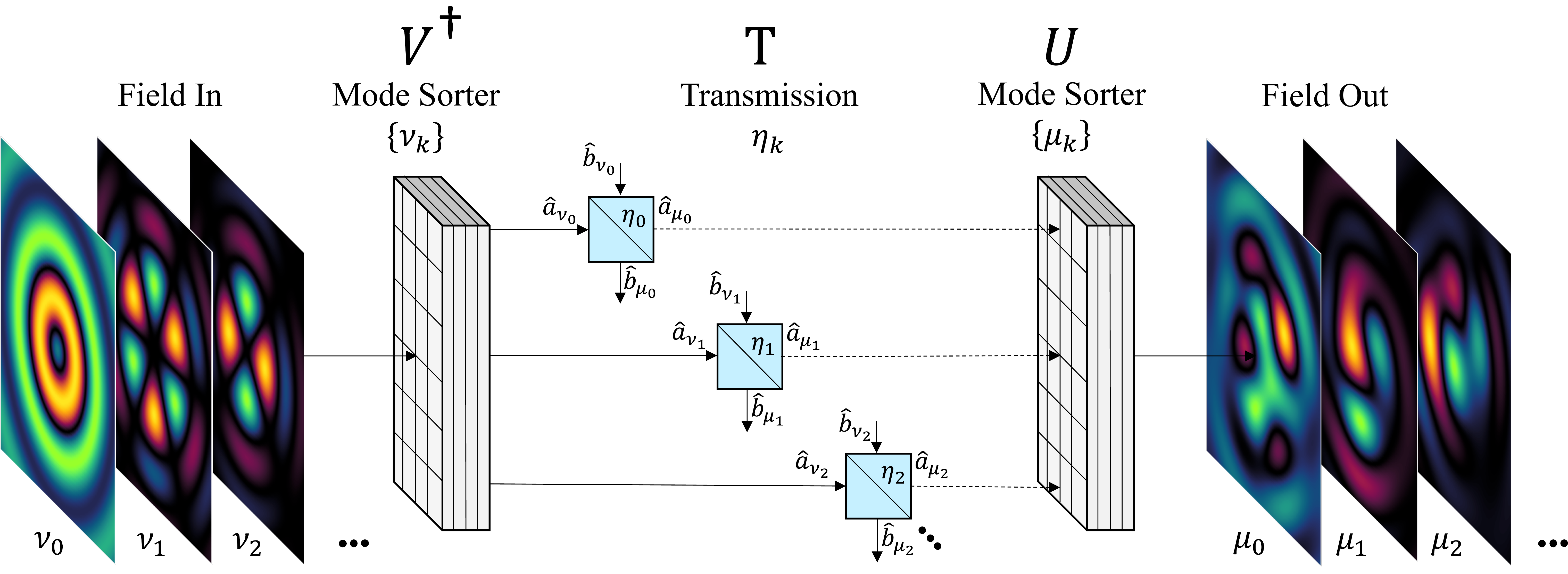}
    \caption{A photonic diagram illustrating the singular value decomposition of an arbitrary coronagraph. The input transverse spatial modes $\hat{a}_{\nu_k}$ are mapped bijectively to the output spatial modes $\hat{a}_{\mu_k}$ with loss $\eta_k$ modeled by a beam-splitter array. We introduce ancillary modes $\hat{b}_{\nu_{k}},\hat{b}_{\mu_{k}}$ to account for energy conservation. One recovers the Kraus operators for the channel by tracing out the ancillary modes.}
    \label{fig: Coronagraph Quantum Channel}
\end{figure*}

Figure \ref{fig: Coronagraph Quantum Channel} provides a photonic model of the coronagraph channel. Two spatial mode sorters, are designed to (de)multiplex spatial modes $\hat{a}_{\nu_k} = \int \nu_{k}^{*}(\mbf{r})\hat{a}(\mbf{r}) d\mbf{r}$ and $ \hat{a}_{\mu_k} = \int \mu_{k}^{*}(\mbf{r})\hat{a}(\mbf{r}) d\mbf{r}$ respectively. The input/output modes couple pair-wise to a beam-splitters parameterized by transmission coefficients $\eta_k$. We introduce an ancillary system via the beam-splitter ports $\{\hat{b}_{\nu_k},\hat{b}_{\mu_k}\}$ to account for lost energy such that the photonic model defines a unitary operation. By injecting vacuum in ports $\hat{b}_{\nu_k}$ and tracing out the ancillary system, one recovers the Kraus operators of Eq. \ref{eqn: Kraus Operators}.

\subsection{Quantum Limits After Propagating Through a Coronagraph Channel}
Given the quantum limits derived for the received state $\hat{\rho}_1$ and the formulation of a coronagraph channel, a natural question arises: what are the quantum limits of exoplanet detection and localization after propagating $\hat{\rho}_1$ through a coronagraph? If the QCE/QRE and QFIM are preserved after propagation, then the coronagraph is equipped to achieve the quantum limits of exoplanet discovery, assuming an optimal measurement is made. In subsequent analysis, we specialize to coronagraphs that completely null the fundamental mode (and potentially attenuate higher order modes) such that $\ket{\nu_0} = \ket{\psi_0}$ with $\eta_0 = 0$. In the high-contrast limit, the single-photon component of the state available after propagation is,
\begin{equation}
    \begin{split}
            \hat{\rho}_1' &= \hat{C}_{1} \hat{\rho}_{1} \hat{C}_{1}^{\dagger} \xrightarrow[]{b<<1} \\
             & b \sum_{k,k'>0}\sqrt{\eta_k \eta_{k'}} \, \Gamma_{\nu_k}(\mbf{r}_{\Delta}) \Gamma^{*}_{\nu_{k'}}(\mbf{r}_{\Delta}) \dyad{\mu_{k}}{\mu_{k'}}
    \end{split}
\end{equation}
Tailoring our information analysis to this surviving single-photon component, the post-propagation QCE and the QFIM are (Appendix \ref{apd: Post-Propagation Quantum Limits}),
\begin{equation}
    \xi_{Q}' \xrightarrow{b<<1} b \sum_{k>0} \eta_{k} |\Gamma_{\nu_k}(\mbf{r}_{\Delta})|^2
    \label{eqn: Post-Propagation QCE}
\end{equation}
\begin{equation}
    \mathcal{K}_{ij}' \xrightarrow{b<<1} 4b \sum_{k>0} \eta_k \Re{\big(\partial_{\theta_i}\Gamma_{\nu_k}(\mbf{r}_{\Delta})\big)^*\big(\partial_{\theta_i}\Gamma_{\nu_{k}}(\mbf{r}_{\Delta})\big)}
    \label{eqn: Post-Propagation QFIM}
\end{equation}
Both post-propagation limits appear as a sum of terms involving the correlation functions of the input modes $\Gamma_{\nu_k}(\mbf{r}_{\Delta})$ weighted by the transmission of each mode. In particular, the post-propagation QCE of Eq. \ref{eqn: Post-Propagation QCE} is indeed the throughput of the coronagraph scaled by the relative brightness of the exoplanet. Thus $\xi_{Q}'$ is the probability that a photon came from the exoplanet and successfully avoided being rejected by the coronagraph. Meanwhile, the post-propagation QFIM of Eq. \ref{eqn: Post-Propagation QFIM} is a sum of classical Fisher information content within each input mode weighted by its transmission.  In general, the post-propagation quantum limits are bounded from above as $\xi_{Q}' \leq \xi_{Q}$ and $\bm{\mathcal{K}}' \leq  \bm{\mathcal{K}}$, where equality occurs when $\eta_{k>0}=1$ such that all modes orthogonal to the fundamental mode are perfectly transmitted.

In appendix \ref{apd: Post-Propagation Quantum Limits}, we show that performing direct imaging and SPADE measurements achieve the post-propagation QCB. Moreover, both measurements also achieve the post-propagation QFI if the SVD of the coronagraph satisfies certain conditions. For direct imaging, the function $\sum_{k} \Gamma_{\nu_k}(\mbf{r}_{\Delta}) \mu_{k}(\mbf{r})$ must be purely real or imaginary. For SPADE, one must measure with the POVM $\dyad{\mu_k}$ and the phase of the correlation coefficients must be parameter-independent such that $\partial_{\theta_i} \arg(\Gamma_{\nu_{k}}(\mbf{r}_{\Delta})) = 0$.


\section{Comparative Survey of Coronagraph Designs}
\begin{figure*}
    \centering
    \includegraphics[width=.9\linewidth]{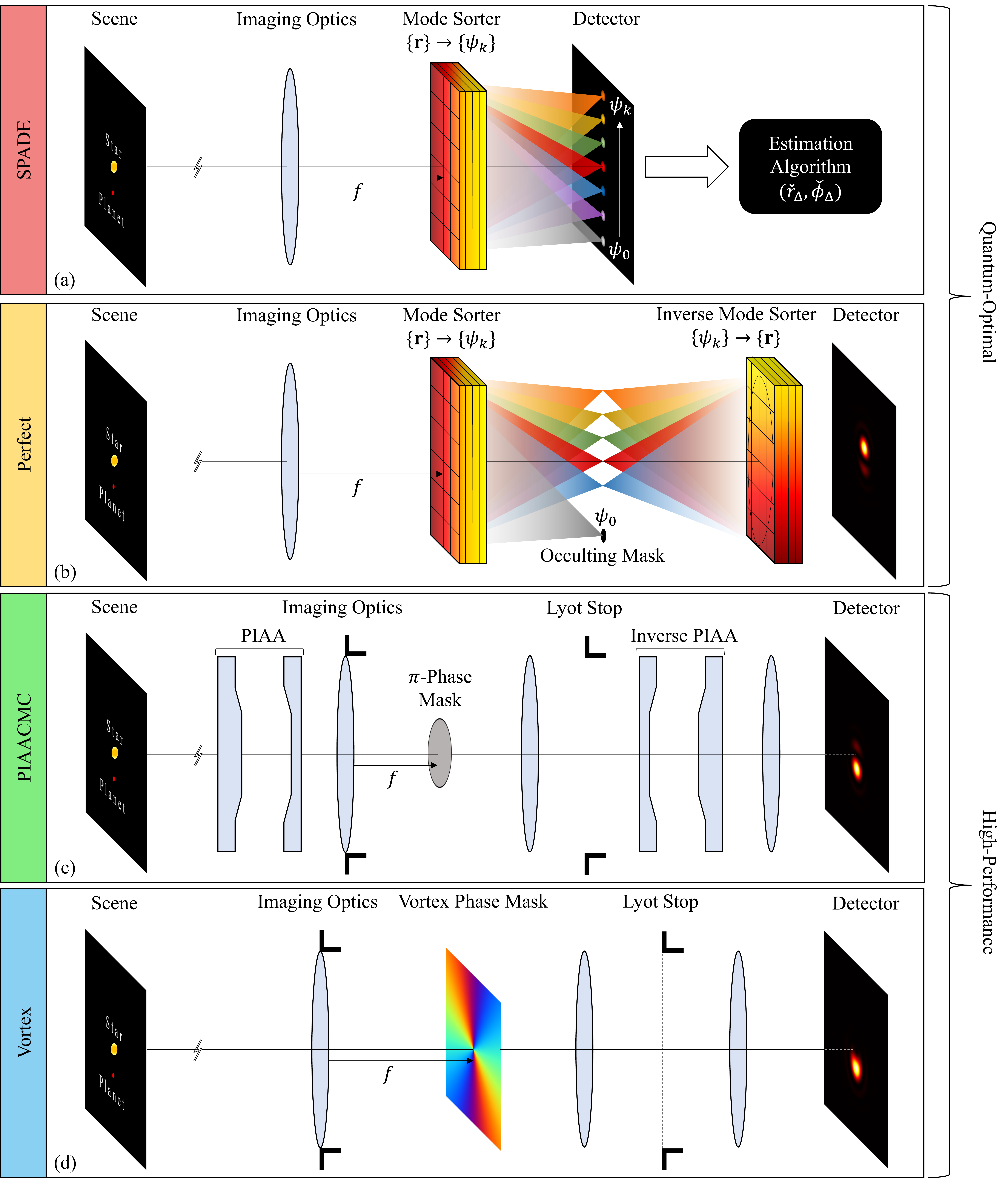}
    \caption{\textbf{(a)} PSF-adapted Spatial Mode Demultiplexing (SPADE): A system that decomposes the incident optical field into a set of PSF-adapted orthonormal transverse spatial modes and counts the number of photons in each mode channel. \textbf{(b)} Perfect Coronagraph (PC): A direct-imaging coronagraph that exclusively eliminates the fundamental mode of the telescope prior to direct detection. \textbf{(c)} Phase-induced Amplitude Apodization Complex Mask Coronagraph (PIAACMC): A direct-imaging coronagraph that employs aspheric optics to induce a lossless prolate spheroidal apodization. The complex phase mask causes the apodized fundamental mode to destructively interfere with itself over the clear aperture of the Lyot stop. \textbf{(d)} Vortex Coronagraph (VC): A direct-imaging coronagraph that diffracts the fundamental mode outside the clear aperture of the Lyot stop using a vortex phase mask with integer topological charge.}
    \label{fig: Coronagraphs}
\end{figure*}

In this section we introduce four unique coronagraph systems (SPADE, PC, PIAACMC, VC) configured to isolate photons in the fundamental mode of a telescope with a circular pupil (see Fig. \ref{fig: Coronagraphs}). In analyzing and comparing these systems, we seek to understand where quantum-optimal coronagraph designs stand to offer valuable performance improvements over existing coronagraphs for exoplanet discovery tasks.

\subsection{Quantum-Optimal Coronagraphs}
\label{sec: QFI-Optimal SPADE}
SPADE and the PC are two systems that saturate the quantum limits of exoplanet detection and localization in the high-contrast regime. Fig. \ref{fig: Coronagraphs}(a) and (b) shows possible implementations of either system using abstract spatial mode sorting elements. Similar implementations of mode-sorter-based coronagraphs have been proposed that employ photonic lanterns \cite{Xin:2022} and multi-plane light converters \cite{Carpenter:2020,Deshler:2025_ExperimentalCoronagraph}.

\subsubsection{PSF-Adapted SPADE}
A PSF-adapted SPADE measurement of the state $\hat{\rho}_1$ is defined through the orthogonal projectors $\{\dyad{\psi_k}: k\in\mathbb{N}\}$ where $\ket{\psi_{0}}$ is the single-photon state of the fundamental mode. Measuring multiple copies of the state amounts to counting the total number of photons arriving in each mode over a given integration time. A positive exoplanet detection is heralded if a photon arrives in any mode other than the PSF mode. An exoplanet is localized by applying an estimator (e.g. maximum likelihood) on the complete set of measurement outcomes. The Fourier-Zernike modes (Appendix \ref{apd: Fourier-Zernike Modes}) constitute one such PSF-adapted basis and are used ubiquitously throughout the numerical simulations underpinning this work.

\subsubsection{Perfect Coronagraph}

The PC is characterized by a direct-imaging system that exclusively removes the fundamental mode from the field prior to detection without affecting higher-order modes. As such, it's Kraus operators are,
\begin{equation}
    \hat{C}_{1} = \hat{I}_{1} - \dyad{\psi_0}  \qquad
    \hat{C}_{0,k} = \delta_{0k} \dyad{0}{\psi_k} 
\end{equation}
where $\hat{I}_{1}$ is the identity on $\mathcal{H}_1$. Fig. \ref{fig: Coronagraphs}(b) depicts one possible implementation of the PC using mode sorters. Like the SPADE system, light is sorted into a set of PSF-adapted modes. However, rather than count photons in each mode channel, the PSF mode channel is rejected with a mask while all other modes are left to propagate freely past the sorting plane. Subsequently, an inverse mode sorter is applied to convert the field in the sorting plane back into an image. 
Thus, a positive exoplanet detection is heralded if a photon arrives at the detector plane at all. An exoplanet is localized by applying an estimator on the irradiance distribution measured at the detector. Unlike SPADE measurements, direct imaging preserves perceptual resemblance to the scene itself which could be useful for a broader variety of astronomical imaging tasks, such as infering the presence of multiple exoplanets and exozodiacal dust disks.

\subsection{High-Performance Coronagraphs}
\label{sec: Coronagraph Survey}

As comparisons to the quantum optimal coronagraphs, we consider two state-of-the-art coronagraphs: the Phase-Induced Amplitude Apodization Complex Mask Coronagraph (PIAACMC) \cite{Guyon:2010} and the Vortex Coronagraph (VC) \cite{Foo:2005}. Both the PIAACMC and VC theoretically achieve complete rejection of the fundamental mode by modulating the phase of the field at an intermediate focal plane using a (generally complex) transmission mask as shown in Fig. \ref{fig: Coronagraphs}(c) and \ref{fig: Coronagraphs}(d). The mask is designed to diffract light in the fundamental mode outside the clear aperture support of a downstream Lyot stop. Finding the closed-form SVDs of the PIAACMC and VC is analytically challenging. Instead we resort to numerical methods to determine their input modes, output modes, and transmissivities. Figure \ref{fig: SVD Modes} shows a selection of the numerically-computed SVD components up to a finite index. For completeness we provide a brief overview of each of these coronagraphs. The interested reader is referred to \cite{Guyon:2006} for a more comprehensive explanation of such advanced Lyot-style coronagraphs. 

\subsubsection{Phase-Induced Amplitude Apodization Complex Mask Coronagraph}

The PIAACMC achieves complete extinction of an on-axis star by applying a pupil apodization in combination with a $\pi$-phase mask placed at the intermediate focal plane of the objective (i.e., Roddier and Roddier Coronagraph \cite{Roddier:1997}). The appropriate spatial profile of the apodization function is a member of the prolate spheroids specific to the geometry of the pupil \cite{Aime:2002, Soummer:2003, Soummer:2005}. These are eigenfunctions of the finite Fourier transform over the aperture support. The PIAACMC employs aspheric optics to redistribute the light over the pupil and achieve the required spatial apodization for complete star extinction \cite{Guyon:2010}. The use of refractive elements to induce apodization as opposed to a spatial transmission mask preserves light throughput past the pupil. Phase modulating the apodized fundamental mode at the focal plane with a $\pi$-phase mask causes the field to destructively interfere with itself such that the amplitude is zero within the clear aperture area of a downstream Lyot stop. For off-axis point sources, self-interference induced by the phase mask does not completely cancel the field amplitude over the clear aperture. Thus, some light successfully propagates past the Lyot stop. Any distortion generated by the first set of aspheric surfaces is compensated by a set of identical but inverted aspheric surfaces before imaging the field onto a detector.

\subsubsection{Vortex Coronagraph}

The vortex coronagraph first proposed by \cite{Foo:2005} introduces a vortex phase plate with integer topological charge at the focal plane of the objective. This element introduces a helical phase modulation of the field. Like the PIAACMC, the phase modulation diffracts light in the fundamental mode outside the clear aperture area of a downstream Lyot stop. Off-axis sources will instead acquire a near-constant phase over the bulk of their field amplitude and pass through the system relatively unaltered. Vortex coronagraphs and vortex fiber nullers have been the subject of several sub-diffraction exoplanet imaging demonstrations in recent years \cite{Mari:2012, Ruane:2018, Ruane:2019, Echeverri:2019, Echeverri:2023, Echeverri:2024}. For all figures and comparative performance analysis presented herein, we consider a charge-2 vortex.

\begin{figure*}
    \centering
    \includegraphics[width=\linewidth]{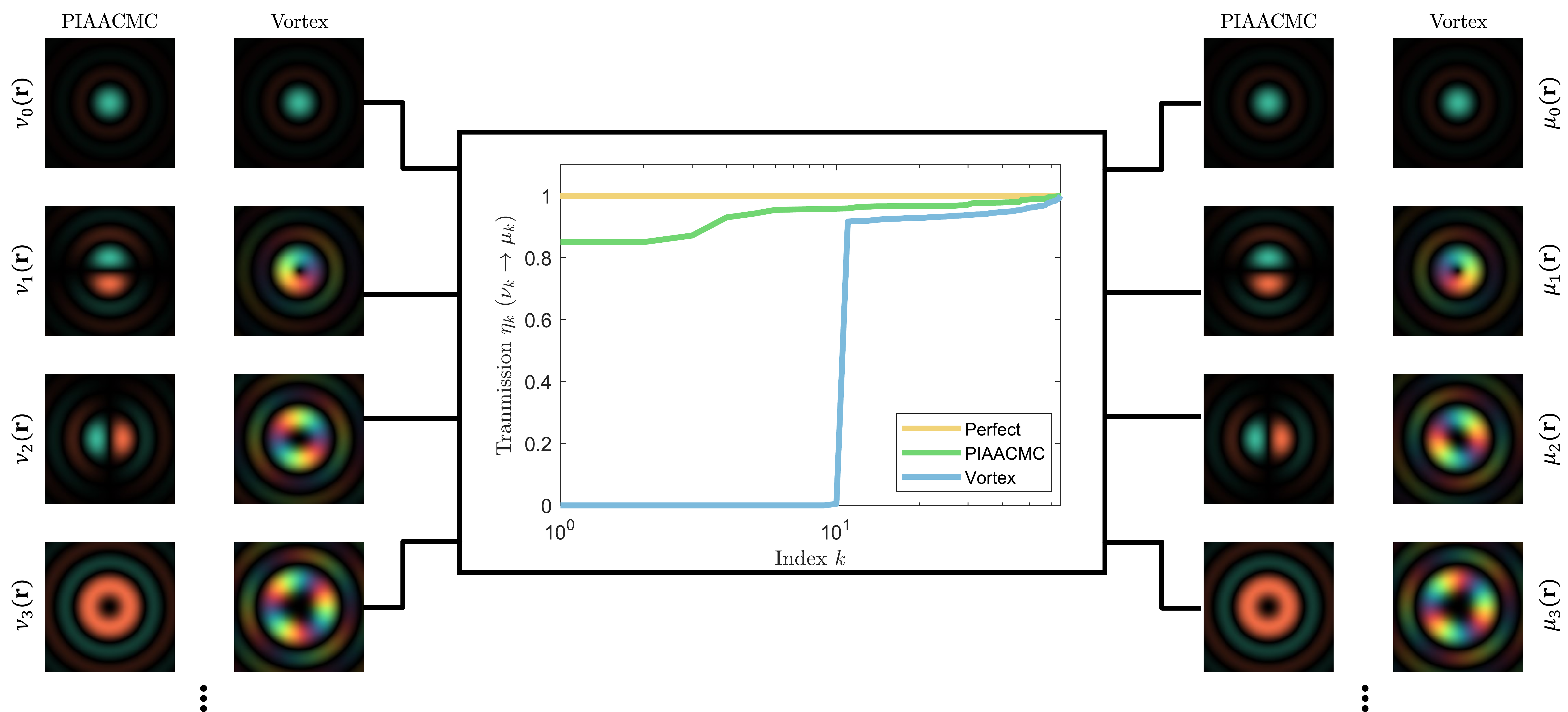}
    \caption{Spatial input and output modes of the PIAACMC and VC ordered by transmission from least to greatest transmission. The mode with lowest transmission in both systems is the PSF of the circular aperture $\nu_0(\mbf{r}) = \psi_0(\mbf{r})$ with transmission coefficient $\eta_0 =0$. The convergence of transmission coefficients for higher order modes $k>0$ are shown in the center plot. }
    \label{fig: SVD Modes}
\end{figure*}

\section{Simulation Results}
\label{sec: Results}

\subsection{Exoplanet Detection Performance Comparison}
\begin{figure*}
    \centering
    \includegraphics[width=\linewidth]{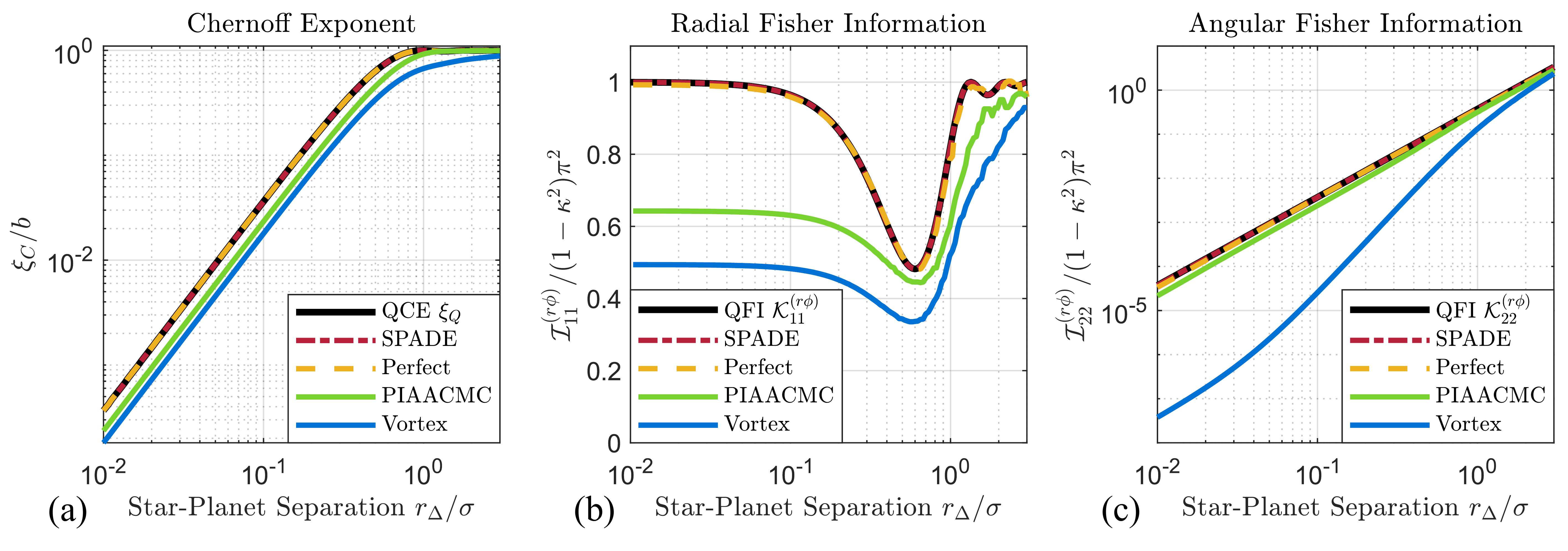}
    \caption{Numerical results for the \textbf{(a)} Chernoff Exponent, \textbf{(b)} Radial Fisher Information, and \textbf{(c)} Angular Fisher Information of each coronagraph as a function of the star-planet separation. The relative brightness was set to $b=10^{-9}$. Quantum optimal systems (SPADE and PC) offer salient improvements over high-performance coronagraphs (PIAACMC and VC) in the sub-diffraction regime.}
    \label{fig: CCE CFI Coronagraphs}
\end{figure*}

In Fig. \ref{fig: CCE CFI Coronagraphs}(a) we compare the classical Chernoff exponent (CCE) of each coronagraph against the QCE at an extreme star-planet contrast. The SPADE and PC saturate the QCE at all contrasts due to complete rejection of the null hypothesis $\hat{\rho}_0$, and minimal attenuation of hypothesis $\hat{\rho}_1$. While the PIAACMC and the VC also completely reject the null hypothesis, they demonstrate sub-optimal performance due to excessive attenuation of higher-order modes that couple light from the exoplanet under hypothesis $\hat{\rho}_1$. The quantum-optimal SPADE and PC systems achieve $\sim 1.6\times$ and $\sim 2\times$ enhancement in the Chernoff exponent over PIAACMC and VC respectively. Under shot-noise-limited conditions, quantum-optimal systems are therefore predicted to reach the same level of exoplanet detection confidence as the PIAACMC and VC with roughly $40\%-50\%$ lower integration time for exoplanets located below the diffraction limit. Table \ref{tab: Integration Times} compares the integration time required for each coronagraph to reach several detection error probabilities.

Fig. \ref{fig: SVD Modes} shows four of the input/ouput modes of the PIAACMC and the VC with lowest transmission factors $\eta_k$. Importantly, some of the transmission coefficients $\eta_{k}$ for $k>0$ are less than one. Those with the lowest transmissivity have non-zero projections onto the tip-tilt modes (i.e. cartesian spatial derivatives of $\psi_0(\mbf{r})$), which are dominantly active for small off-axis displacements of the exoplanet. The superior performance of the PIAACMC compared to the VC at small star-planet separations stems from its comparatively higher transmission of tip-tilt modes. 

\begin{table*}
    \centering
    \begin{tabular}{c | c c c}
               &  Perfect/SPADE & PIAACMC & Vortex \\
         \hline 
         $P_e$ &  & \textbf{Detection Integration Times (seconds)}   &\\
         \hline
         $10^{-1}$ & 1,073 & 1,663 & 2,167 \\
         $10^{-2}$ & 2,146 & 3,326 & 4,334 \\
         $10^{-3}$ & 3,220 & 4,990 & 6,501 \\
         $10^{-4}$ & 4,293 & 6,653 & 8,668 \\
         \hline
         $\sigma_{loc}/r_{\Delta}$ &  & \textbf{Localization Integration Times (seconds)}  & \\
         \hline
         $1$ & 6 & 9 & 33\\
         $1/2$ & 25& 37 & 134 \\
         $1/10$ & 626& 927 & 3,350 \\
         $1/100$ & 62,596 & 92,749 & 335,000 \\
    \end{tabular}
    \caption{Integration times required to reach particular detection error probabilities $P_e$ and relative localization errors $\sigma_{loc}/r_{\Delta}$ with different coronagraphs. These times are calculated for an example star-planet system with $r_{\Delta}/\sigma = 0.1$ and $b=10^{-9}$. The star photon flux and telescope specifications used are given in Table \ref{tab: Reference System for Integration Times}.}
    \label{tab: Integration Times}
\end{table*}

\subsection{Exoplanet Localization Performance Comparison}

In Fig. \ref{fig: CCE CFI Coronagraphs}(b) and \ref{fig: CCE CFI Coronagraphs}(c) we directly compare the radial and angular classical Fisher information (CFI) of SPADE, PC, PIAACMC and VC against the QFI at an extreme star-planet contrast. When the star and exoplanet are well-separated $(r_{\Delta} \gtrsim 2\sigma)$, then the CFI for all systems converge to the quantum bound, as light begins coupling to higher-order modes that experience minimal attenuation. However, at sub-diffraction star-planet separations $(r_{\Delta}<\sigma)$, only SPADE and PC fully saturate the QFI. Furthermore, the SPADE and PC information curves are strictly greater than those of the PIAACMC and VC. For the radial parameter $r_{\Delta}$, Fig. \ref{fig: CCE CFI Coronagraphs}(b) indicates that SPADE and PC are roughly $2\times$ as information-efficient as PIAACMC and VC in the deeply sub-diffraction regime $(r_{\Delta} \lesssim \sigma/10)$, thereby reducing the localization error by a factor of $\sqrt{2}$ for the same photon budget. This performance difference arises due to lower-order mode attenuation in the PIAACMC and VC as shown in \ref{fig: SVD Modes}(b).

\begin{figure*}
    \centering
    \includegraphics[width=\linewidth]{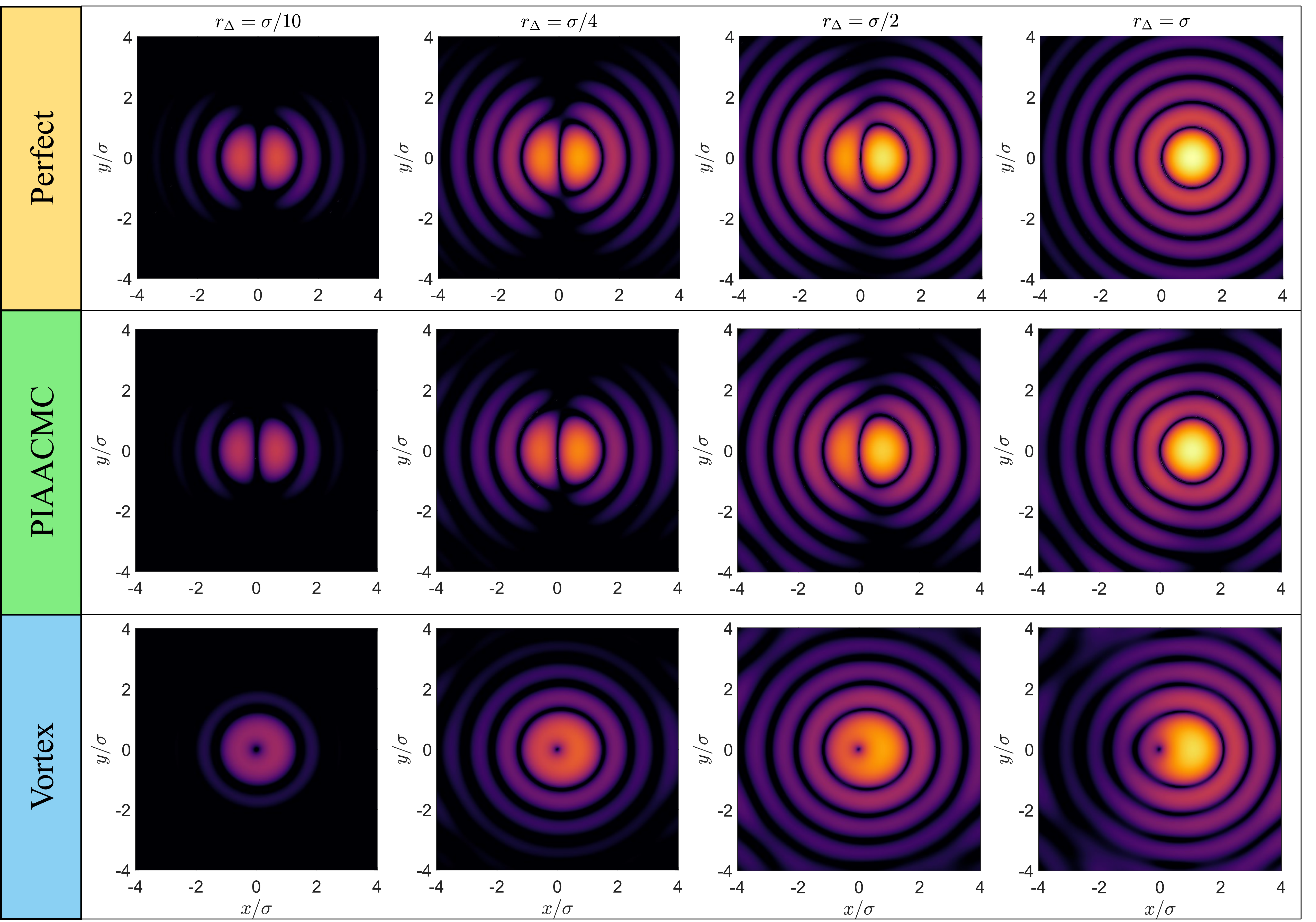}
    \caption{Theoretical intensity distributions (log-scale) for direct-imaging coronagraphs pointed at the center of intensity of a star-planet system with $b=10^{-9}$. Each column corresponds to a different star-planet separation along the x-axis. The two-lobe structure visible in the sub-diffraction regime (column 1) for the PC and PIAACMC emerges from dominant coupling to the tip-tilt modes at small separations. Note that the vortex coronagraph does not exhibit a two-lobe structure as the tilt modes are highly attenuated. Consequently, the sensitivity of the Vortex coronagraph to radial separation $r_{\Delta}$ and the angular orientation of the star-planet system suffers as shown by the sub-optimal CFI curves of Fig. \ref{fig: CCE CFI Coronagraphs}.}
    \label{fig: Direct Imaging Coronagraphs}
\end{figure*}

For the angular parameter $\phi_{\Delta}$, Fig. \ref{fig: CCE CFI Coronagraphs}(C) shows that both SPADE and PC saturate the QFI limit. In the sub-diffraction regime, SPADE and the PC are approximately $ 2\times$ and $100 \times$ more information efficient than the PIAACMC and VC respectively. The drastic sub-optimality of the VC at small star-planet separations can be understood by inspecting simulated intensity images for each coronagraph shown Fig \ref{fig: Direct Imaging Coronagraphs}. The PC images exhibits a double-lobed structure characteristic of strongly-coupled tip-tilt modes after removal of the fundamental mode. The PIAACMC images exhibit nearly identical structure as those of the PC with notably lower transmission, implying that the the PIAACMC nulls the fundamental mode and small amounts of the tip-tilt modes. Conversely, the VC creates doughnut-like structures at sub-diffraction separations which increase uncertainty about the angular orientation of the star-planet axis. 

\begin{figure*}
    \centering
    \includegraphics[width =\linewidth]
    {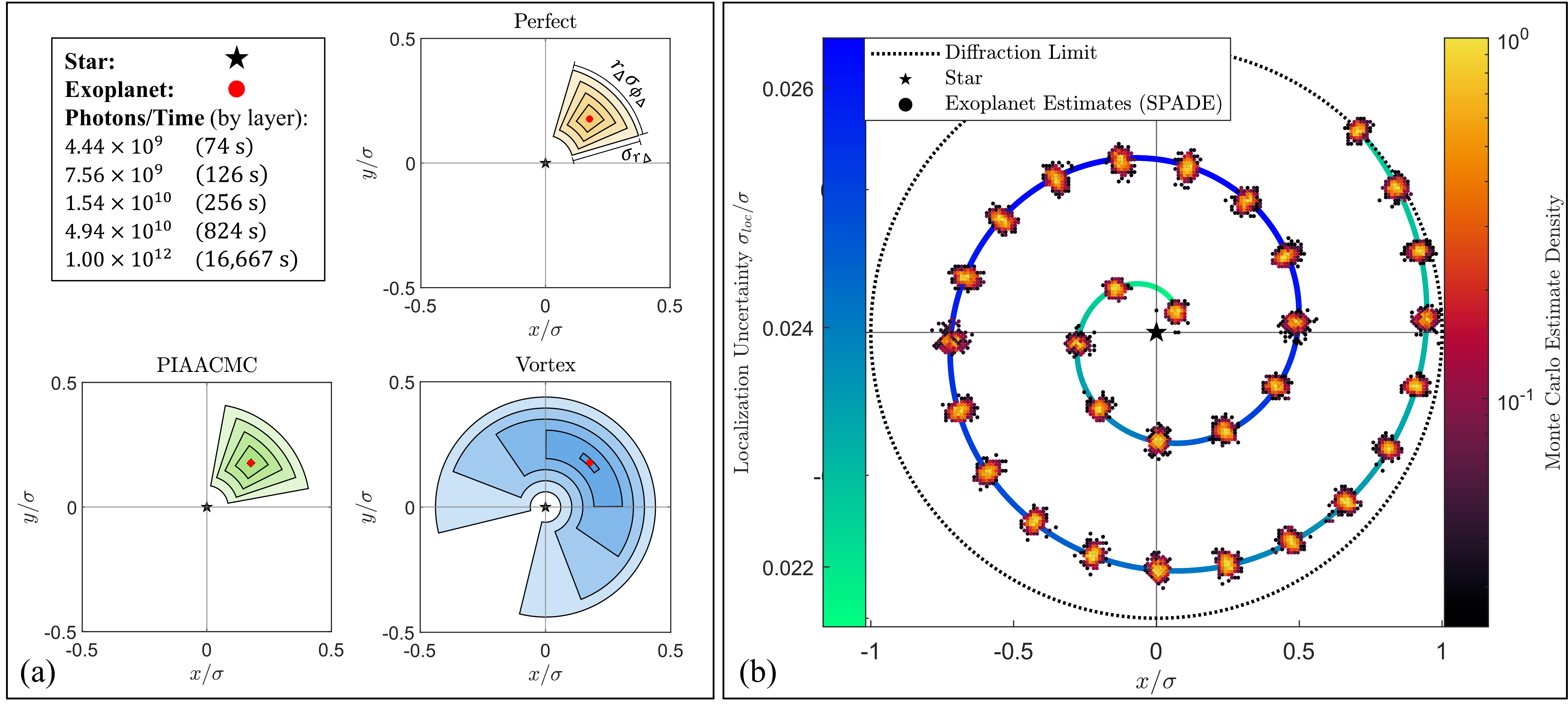}
    \caption{\textbf{(a)} Quantified variation in the size of the uncertainty patch for each cluster of exoplanet estimates. For each cluster, we fit a 2D Gaussian distribution and compute an uncertainty patch from the covariance matrix. We see that the size of the uncertainty patches roughly track in accordance with the fundamental bound on the localization uncertainty $\sigma_{loc}$.\textbf{(b)} Monte-Carlo simulations showing localization of an exoplanet at various sub-Rayleigh distances from the star location using maximum likelihood estimation on SPADE measurements in the Fourier Zernike basis, truncated to maximum radial order $\text{max}(n) = 10$. For each exoplanet location, $500$ measurement and estimation trials were simulated. The brightness contrast was set to $b=10^{-9}$ and the mean photon count per measurement was taken to be $300\times 10^{9}$ such that, on average, $\sim 300$ photons from the exoplanet were collected per measurement. We have chosen exoplanet locations spaced equally over a spiral to illustrate variations in uncertainty that may depend on radial or angular position in the field of view.}
    \label{fig: FZ Modes and Monte-Carlo Estimates}
\end{figure*}

The relative difference in Fisher information of each system translates to different exposure times required to achieve a particular exoplanet localization precision, which is of practical significance. Table \ref{tab: Integration Times} provides a ready comparison of these exposure times between each coronagraph. In general, we observe that that SPADE and PC offer significant performance improvements over PIAACMC and VC primarily in the sub-diffraction regime, in terms of $1.5 \times$ to $2 \times$ lower exposure times respectively. To illustrate the prospective value of quantum-optimal systems, Fig. \ref{fig: FZ Modes and Monte-Carlo Estimates}(a) visually depicts the size of the localization uncertainty patch associated with each coronagraph for different numbers of photons allotted to localizing a sub-diffraction exoplanet. Fig. \ref{fig: FZ Modes and Monte-Carlo Estimates}(b) depicts Monte-Carlo simulations of a SPADE system localizing an exoplanet positioned about various locations in sub-diffraction regime. In particular, we apply maximum likelihood estimates to the SPADE measurements in order to recover the exoplanet positions. The distribution of maximum likelihood estimates are strongly concentrated around the ground truth positions of the exoplanet.

\section{Discussion}
\label{sec: Discussion}

In review, we report the fundamental quantum bounds of exoplanet detection (QCE, QRE) and localization (QFIM). We also identify intuitive interpretations of these bounds in the high-contrast limit. In the case of exoplanet detection, the QCE and QRE are simply the probability that a photon originates from the exoplanet and is detected outside the fundamental mode. Meanwhile, the QFI can be thought of as the constant information available for localizing a single point source (i.e., the exoplanet) minus the classical Fisher information supplied by the fundamental mode. We present two quantum-optimal coronagraphs, SPADE and the Perfect Coronagraph, which saturate the QCE and QFI bounds. As points of comparison, we numerically show that two competitive coronagraphs, the PIAACMC and the Vortex Coronagraph, are sub-optimal in the regime of small star-planet separations due to  excess attenuation of higher-order modes beyond the fundamental mode of the telescope.

Encouraged by these results, several future research directions appear ripe for exploration. On a theoretical front, a QFIM analysis remains to be done involving simultaneous estimation of the star-planet separation $\mbf{r}_{\Delta}$, the centroid $\mbf{R}$, and the relative brightness $b$ in 2D. The centroid and brightness parameters are considered contextual nuisance parameters. Nevertheless, imperfect knowledge of these nuisance parameters may ultimately affect the imprecision limits of the star-planet separation $\mbf{r}_{\Delta}$. In \cite{Rehacek:2017}, the QFIM for simultaneous estimation of the centroid, relative brightness, and separation is derived in 1D. The authors also propose a numerically-optimized measurement for simultaneously saturating the QCRB. Meanwhile, in \cite{Grace:2020} the authors posit an optimal multi-stage receiver where the initial stages are dedicated to collecting sufficient prior information on nuisance parameters so as to ensure the primary estimation task is executed optimally under constrained photon resources. It would also be valuable to perform a QFIM analysis for a resolved star with finite extent. This analysis would help determine whether classical propositions for even-order coronagraphs \cite{Guyon:2006,Belikov:2021} are optimal strategies.    

Analysis quantifying the robustness of quantum-optimal instruments in the context of real-world imperfections will also be required to assess their prospective utility as an emerging technology. In particular, we expect low-order wavefront instability \cite{Ruane:2018}, finite spectral bandwidths, and detector noise to present the most salient challenges for optimal direct imaging coronagraphs. Meanwhile, for SPADE devices we expect modal cross-talk to also hamper performance. Fortunately, related analysis is already underway. 

In the context of exoplanet detection, \cite{Schlichtholz:2024} showed that any amount of modal cross-talk in SPADE devices immediately demands introducing \textit{a priori} knowledge of the exoplanet position in order to execute a maximum likelihood hypothesis test. Later, \cite{Linowski:2025_ExoplanetQuantumLimitsNoise} evaluated the impact of cross-talk and dark current on the performance of SPADE and PC showing that these sources of error cause departure from the quantum Stein lemma at similar rates. Here, the author also formulate a decision strategy for the hypothesis test that accounts for starlight leakage induced by modal cross-talk.
 
In the context of exoplanet localization, \cite{Prasad:2023_UniformDisk} recently reported the QFI for the position of an exoplanet against a uniformly bright disk. Such analysis targets the quantum limits of transit photometry. In yet earlier work, \cite{Prasad:2022_Broadband} derived the QFI for the two-source separation with a general spectral power density, while \cite{Grace:2022_Thesis} discovered an important trade-space while performing a Fisher information analysis of SPADE in the context of finite spectral bandwidths. As the authors show, a broader spectral bandwidth increases photon flux at the expense of greater mode mismatch. Thus, the optimal spectral band-pass filter for a SPADE receiver scales as $\frac{\Delta \lambda }{\lambda_0} \propto \frac{r_{\Delta}}{\sigma}$. In a related vein, numerical demonstrations of simultaneous spectral and spatial mode sorting using multi-plane light conversion have recently been reported \cite{Zhang:2020}. Such techniques may enable attainment of quantum-optimal exoplanet detection/localization while concurrently performing spectroscopy.

In this work, we have considered a scene consisting of a star and a single exoplanet. However, it is conceivable that a star is surrounded by multiple exoplanets. In this setting, a direct-imaging coronagraph like the PC would appear to be favorable as it provides a measurement that bears perceptual resemblance to the scene itself (minus the star). However, if the pair-wise separation between multiple exoplanets is below the diffraction limit, then spatial mode sorting measurements are likely to provide better localization estimates. A poignant example is the adaptive Bayesian SPADE technique developed in \cite{KitLee:2023} for localizing multiple sub-diffraction point sources and estimating their brightness. This technique substantially outperforms direct imaging and could be applied to coronagraphy by adaptively updating the mode basis spanning the orthogonal complement space of the fundamental mode. Adaptive mode sorting techniques may also enable integration between wavefront error correction and SPADE measurements as recently suggested in \cite{Trzaska:2024_PhaseRetrieval,Trzaska:2024_Wavefront}. 

The quantum limits of exoplanet detection and localization can be achieved simultaneously by separating the fundamental mode of the imaging system without attenuating orthogonal modes in the process. The astronomy community has long intuited this to be an optimal strategy. By conducting a quantum information analysis, we ground this intuition on stronger theoretical footing. In particular, we introduce a quantum channel formalism which provides a systematic mathematical treatment of any coronagraph and facilitates comparative analyses. Finally, we find that the QFI for localization persists deep into the sub-diffraction regime where many undiscovered exoplanets are suspected to exist. The quantum-optimal instruments discussed in this work may therefore expand the domain of accessible exoplanets.

\begin{acknowledgements}
The authors would like to thank Dr. Aqil Sajjad, Dr. Saikat Guha, Dr. Daniel Soh, and Itay Ozer for their insightful supporting discussions and contributions.
\end{acknowledgements}

\begin{contribution}
    ND wrote the manuscript, formulated the quantum channel descriptions of coronagraphs, and wrote the code-base for numerical analysis and figure generation. SH and AA both supported writing of the manuscript and provided expertise on coronograph modeling and quantum information analyses.
\end{contribution}

\section*{Funding}
This material is based upon work supported by the National Science Foundation Graduate Research Fellowship under Grant No. DGE-2137419.

\section*{Conflict of Interest}
The authors declare no conflicts of interest.

\section*{Data Availability}
A code-base for (1) generating the figures presented in this work (2) simulating field propagation through the PC, PIAACM, and VC (3) determining coronagraph operators, and (4) simulating exoplanet localization with SPADE can be found in the project GitHub Repository \cite{Deshler:2024_Exoplanet_Quantum_Limits_Github}.

\bibliographystyle{aasjournalv7}
\bibliography{IEEEabrv,references}

\appendix

A critical quantity throughout all analyses in the appendix is the overlap integral between the single photon states of the shifted PSF modes excited by the star and the exoplanet 
\begin{equation}
\Delta \equiv \braket{\psi(\mbf{r}_s)}{\psi(\mbf{r}_e)} = \Gamma_{\psi_0}(\mbf{r}_{\Delta})
\label{eqn: Overlap Parameter}
\end{equation}
where the second inequality arises from the shift-invariant property of the PSF. Unless specified otherwise, our only assumption is that $\psi_0(\mbf{r})$ is a real-valued PSF such that $\Delta \in \mathbb{R}$.
\vspace{2ex}

\section{Towards Helstrom Bound on Exoplanet Hypothesis Testing}
\label{apd: Towards Helstrom Bound}
In this section we consider the quantum limits of exoplanet detection given by the Helstrom bound where joint measurements are allowed across multi-copy states. We will specialize to the discrimination of single photon states $\hat{\rho}_{0}^{\otimes N}$ and $\hat{\rho}_{0}^{\otimes M}$ (as opposed to $\hat{\tilde{\rho}}_{0}^{\otimes M}$ and $\hat{\tilde{\rho}}_{1}^{\otimes M}$) which practically amounts to storing only those states for which a photon interacted with a quantum register - a more general derivation would consider the vacuum contributions as well. Moreover, we will make the simplifying assumption that we are working in the high-contrast regime wherein $\ket{\psi(\mbf{r}_s)}\approx \ket{\psi_0}$ and $\ket{\psi_{\mbf{r}_e}} \approx \ket{\psi(\mbf{r}_\Delta)}$. For notational convenience we will assign $\ket{\underbar{0}}=\ket{\psi_0}$ and $\ket{\underbar{1}}\equiv \ket{\psi(\mbf{r}_{\Delta})}$. Here, $\ket{\underbar{0}}$ is not to be confused with the vacuum state from the main text but is instead a single-photon excitation of the fundamental mode while $\ket{\underbar{1}}$ is a single-photon state of the shifted PSF mode associated with the position of the exoplanet.

The span of these vectors forms a two-dimensional Hilbert space $\underline{\mathcal{H}} \equiv \text{span}(\ket{\underbar{0}},\ket{\underbar{1}}) \subset \mathcal{H}$. We now consider the tensor product space $\underline{\mathcal{H}}^{\otimes N}$ which represents the joint system of $N$ photon arrivals - one photon per temporal coherence interval. Each temporal coherence interval can be thought of as its own independent quantum system such that the space $\underline{\mathcal{H}}^{\otimes N}$ is the product space of multiple systems. Under this formulation, the tensor product density operators are given by,
\begin{align*}
    \hat{\rho}_0^{\otimes N} &= \dyad{\mbf{0}} , \qquad \ket{\mbf{0}} = \underbrace{\ket{\underbar{0},\ldots,\underbar{0}}}_{N}\\
    \hat{\rho}_1^{\otimes N} &= \sum_{\mbf{x}\in \mathcal{X}}p_0^{n_0(\mbf{x})}p_1^{n_1(\mbf{x})}\dyad{\mbf{x}}, \qquad p_0 = 1-b, \,\,  p_1 = b
\end{align*}
where $\mathcal{X}=\{\underbar{0},\underbar{1}\}^{N}$ is the Hamming space of binary codes of length $N$. One instance of such a code might be $\ket{\mbf{x}} = \ket{\underbar{0},\underbar{1},\underbar{1},\ldots,\underbar{0}}$. Additionally, $n_0(\mbf{x})$ and $n_1(\mbf{x})$ are the number of $\underbar{0}$'s and $\underbar{1}$'s in $\mbf{x}$ respectively. Physically, each $\mbf{x}$ represents a possible sequence of photon arrivals where a $\underbar{0}$ or $\underbar{1}$ distinguishes which mode each photon arrived in. Note that the inner product between any pair of joint states $\ket{\mbf{x}},\ket{\mbf{y}}$ for $\mbf{x},\mbf{y}\in\mathcal{X}$ is given by,
$$
\braket{\mbf{x}}{\mbf{y}} = \Delta^{d(\mbf{x},\mbf{y})},\qquad d(\mbf{x},\mbf{y}) =\texttt{sum} [\mbf{x} \oplus \mbf{y}] \in [0:N]
$$
where $d(\mbf{x},\mbf{y})$ is the Hamming distance between two binary codes defined with $\oplus$ representing the XOR logical operation. That is, $d(\mbf{x},\mbf{y})$ is simply the number of places where the vectors $\mbf{x}$ and $\mbf{y}$ differ. The Helstrom bound on the minimum probability of error is given by,
$$
P_{e,min} = \frac{1}{2}\bigg[1 - || q_1 \hat{\rho}_1^{\otimes N} - q_0 \hat{\rho}_0^{\otimes N}|| \bigg] , \qquad ||\hat{A}|| \equiv \Tr |\hat{A}|,\qquad |\hat{A}|\equiv \sqrt{\hat{A}^{\dagger}\hat{A}}
$$
Computing the Helstrom bound numerically runs quickly into a curse of dimensionality as the dimension of $\underline{\mathcal{H}}^{\otimes N}$ grows as $2^N$. To make matters worse, in high-contrast imaging contexts it is expected that the number of photons we must consider is on the order $N \geq 10^9$ to detect Earth-like faint exoplanets with $b\leq 10^{-9}$. Here we do not find a closed-form expression for the Helstrom bound, but we do determine upper and lower bounds of the Helstrom limit itself and provide a simple joint measurement that achieves identical asymptotic error scaling. At worst, this joint measurement is within a factor of two from the Helstrom Bound.

\subsection{Bounds on the Helstrom Limit}

Consider the trace distance between two quantum states $\hat{\rho}$ and $\hat{\sigma}$ given by
$$
D_{tr}(\hat{\rho},\hat{\sigma}) = \frac{1}{2}\Tr|\hat{\rho}  - \hat{\sigma}|
$$
The trace distance was found to be bounded from below in \cite{Puchala_Miszczak:2009_TraceDistBound} via the quantum superfidelity,
$$
G(\hat{\rho},\hat{\sigma})^2 = \Tr[\hat{\rho}\hat{\sigma}] + \sqrt{1-\Tr[\hat{\rho}^2]}\sqrt{1-\Tr[\hat{\sigma}^2]}
$$
and from above via the quantum fidelity \cite{Nielsen_Chuang:2010_QuantumInformation} 
$$
F(\hat{\rho},\hat{\sigma})^2 = \bigg(\Tr[\sqrt{\hat{\rho}^{\frac{1}{2}}\hat{\sigma}\hat{\rho}^{\frac{1}{2}}}]\bigg)^2$$
in the following way:
$$
1 - G(\hat{\rho},\hat{\sigma})^2 \leq D_{tr}(\hat{\rho},\hat{\sigma})\leq \sqrt{1-F(\hat{\rho},\hat{\sigma})^2}
$$
In the case where one of the states is a pure state $\hat{\rho} = \dyad{\psi}$, the superfidelity and the fidelity are equal such that the lower-bound on the trace-distance becomes,
$$
1 - F(\ket{\psi},\hat{\sigma})^2 \leq D_{tr}(\ket{\psi},\hat{\sigma})\leq \sqrt{1-F(\ket{\psi},\hat{\sigma})^2}
$$
In the context of our hypothesis testing the fidelity between our two hypotheses is,
\begin{align*}
F(\hat{\rho}_0^{\otimes N},\hat{\rho}_1^{\otimes N})^2 = F(\ket{\mbf{0}},\hat{\rho}_{1}^{\otimes N})^2 &=\mel{\mbf{0}}{\bigg[{\sum_{\mbf{x}\in\mathcal{X}}}p_0^{n_0(\mbf{x})}p_1^{n_1(\mbf{x})}\dyad{\mbf{x}}\bigg]}{\mbf{0}}\\
&= \sum_{\mbf{x}\in\mathcal{X}} p_0^{n_0(\mbf{x})}p_1^{n_1(\mbf{x})} (|\Delta|^2)^{d(\mbf{0},\mbf{x})}\\
&= \sum_{n=0}^{N} \binom{N}{n} p_{0}^{N-n} p_{1}^{n} (|\Delta^2|)^{n}
\end{align*}
The last line is the expectation for a function $(|\Delta|^2)^{n}$ of a binomial random variable $n$. Writing $(|\Delta|^2)^{n} = e^{t n}$ (with $t \equiv \ln |\Delta|^2$) we find that squared quantum fidelity evaluates to the moment generating function of the binomial distribution:
\begin{align*}
F(\hat{\rho}_0^{\otimes N},\hat{\rho}_1^{\otimes N})^2 = \langle e^{tn}\rangle_{\text{Binom}(n;p_1,N)} =  (p_0 + p_1 e^{t})^{N} = [(1-b) + b|\Delta|^2 ]^N = [1-\xi_{Q}]^N
\end{align*}
where $\xi_Q = b(1- |\Delta^2|) = b(1-p_{\psi_0}(\mbf{r}_\Delta))$ as previously found in Equation \ref{eqn: High-Contrast QCE}. In summary, the fidelity between the joint states in our hypothesis test is given by,

\begin{equation}
    \boxed{
    F^2 \equiv F(\hat{\rho}_0^{\otimes N},\hat{\rho}_1^{\otimes N})^2=  [1-\xi_{Q}]^{N} 
    , \qquad \xi_{Q} = b[1-p_{\psi_0}(\mbf{r}_{\Delta})]}
    \label{eqn: Joint State Fidelity}
\end{equation}
Since $\xi_Q$ is the probability that a single photon is observed outside the fundamental mode, the squared fidelity is equal to the probability that all photons appear in the fundamental mode.  Our goal now is to use this result to establish upper and lower bounds on the Helstrom limit. We will specialize to the case where the priors $q_0 = q_1 = \frac{1}{2}$ are balanced as this appears to be a reasonable assumption for the distribution of Earth-like exoplanets around Sun-like stars (see introduction). Then the Helstrom bound reduces to:

$$
P_{e,min} = \frac{1}{2}\bigg[1-\frac{1}{2}||\hat{\rho}_1^{\otimes N }-\hat{\rho}_0^{\otimes N }||\bigg] = \frac{1}{2}\bigg[1-D(\hat{\rho}_0^{\otimes N },\hat{\rho}_1^{\otimes N})\bigg]
$$
Therefore,
$$
\frac{1}{2}\bigg[ 1 - \sqrt{1-F^2}\bigg]\leq P_{e,min} \leq \frac{1}{2}F^2
$$
In the asymptotic limit of many copies $N\rightarrow \infty$ we Taylor expand the lower bound to arrive at,
$$
\frac{1}{4}F^2 \leq P_{e,min} \leq \frac{1}{2} F^2 \qquad \text{as} \qquad N\rightarrow \infty
$$
Whether or not the lower bound is asymptotically tight remains an open question. Next, consider the simple binary joint measurement $\hat{\Pi}_0 \equiv \dyad{\mbf{0}}$ and $\hat{\Pi}_1 =\hat{I} - \hat{\Pi}_0$
where $\hat{I}$ is the identity on $\mathcal{H}^{\otimes N}$. It is straightforward to show that the probability of detection error under balanced priors ($q_0 = q_1 = \frac{1}{2}$) is given by,
\begin{align*}
P_{e} = \frac{1}{2} \bigg[\Tr[\hat{\Pi}_{0} \hat{\rho}_{1}^{\otimes N}] +\Tr[\hat{\Pi}_{1} \hat{\rho}_{0}^{\otimes N}]   \bigg] = \frac{1}{2} \mel{\mbf{0}}{\hat{\rho}_{1}^{\otimes N}}{\mbf{0}} = \frac{1}{2} F^2\\
\end{align*}
Therefore, the probability of error for this joint measurement is, at worst, a factor of two above the optimal joint measurement on multi-copy states in the asymptotic limit on $N$. Fortunately, this joint measurement can be implemented as a copy-by-copy measurement where one continuously monitors the number of photons arriving in the fundamental mode across all $N$ temporal coherence intervals.

\section{Derivation of the Quantum Chernoff Exponent}
\label{apd: Proof of High-Contrast QCE}

 The single-photon QCE for discriminating between a single point source located on-axis and an arbitrary normalized incoherent exitance distribution $m(\mathbf{r}) \geq 0$ is given by \cite{Grace:2022},

$$
\xi_{Q} = -\ln \bigg( \int m(\mathbf{r}) |\Gamma_{\psi_0}(\mathbf{r})|^2 d^2\mathbf{r} \bigg) 
$$

Moreover, the authors prove that this bound is achieved using a binary spatial mode sorter which sorts the PSF mode $\hat{\Pi}_{\psi_0} = \dyad{\psi_0}$ and its orthogonal complement $\hat{\Pi}_{\perp} =   \hat{I}_{1} - \hat{\Pi}_{\psi_0}$. In our star-planet model we have the following exitance distribution,

$$
m(\mathbf{r}) = (1-b)\delta(\mathbf{r} + b\mathbf{r}_{\Delta}) + b\delta(\mathbf{r} - (1-b)\mathbf{r}_{\Delta})
$$

Inserting this into the QCE expression, we resolve Eq. \ref{eqn: QCE Single Photon}. 

$$
\xi_{Q}(\mathbf{r}_{\Delta}) = - \ln \bigg[ (1-b) |\Gamma_{\psi_0}(-b\mathbf{r}_{\Delta})|^2 + b |\Gamma_{\psi_0}((1-b)\mathbf{r}_{\Delta})|^2 \bigg] 
$$

To determine the high-contrast Quantum Chernoff Exponent, we Taylor expand Eq. \ref{eqn: QCE Single Photon} to first order in $b$. Note that $p_0(\mbf{r})=|\Gamma_{\psi_0}(\mbf{r})|^2$. Moreover, we invoke the property $p_0(0) = 1$, since all light from emitted by an on-axis point source couples to the fundamental mode, and make use of the fact that $\partial_{\mbf{r}}p_0(\mbf{r})|_{\mbf{r}=0} = 0$ which follows directly from the recognition that $p_0(0)$ is necessarily a local maxima. Applying these to the Taylor expansion, it can be shown that,

$$
\xi_{Q}(\mathbf{r}_{\Delta}) \approx  b( 1 - p_0(\mbf{r}_{\Delta})) = b( 1 - |\Gamma_0(\mbf{r}_{\Delta})|^2)
$$

which is the high-contrast QCE of Eq. \ref{eqn: High-Contrast QCE}.

\section{Derivation of the Quantum Stein Lemma and Quantum Relative Entropy}
\label{apd: Quantum Stein Lemma}
\begin{figure}
    \centering
    \includegraphics[width=.5\linewidth]{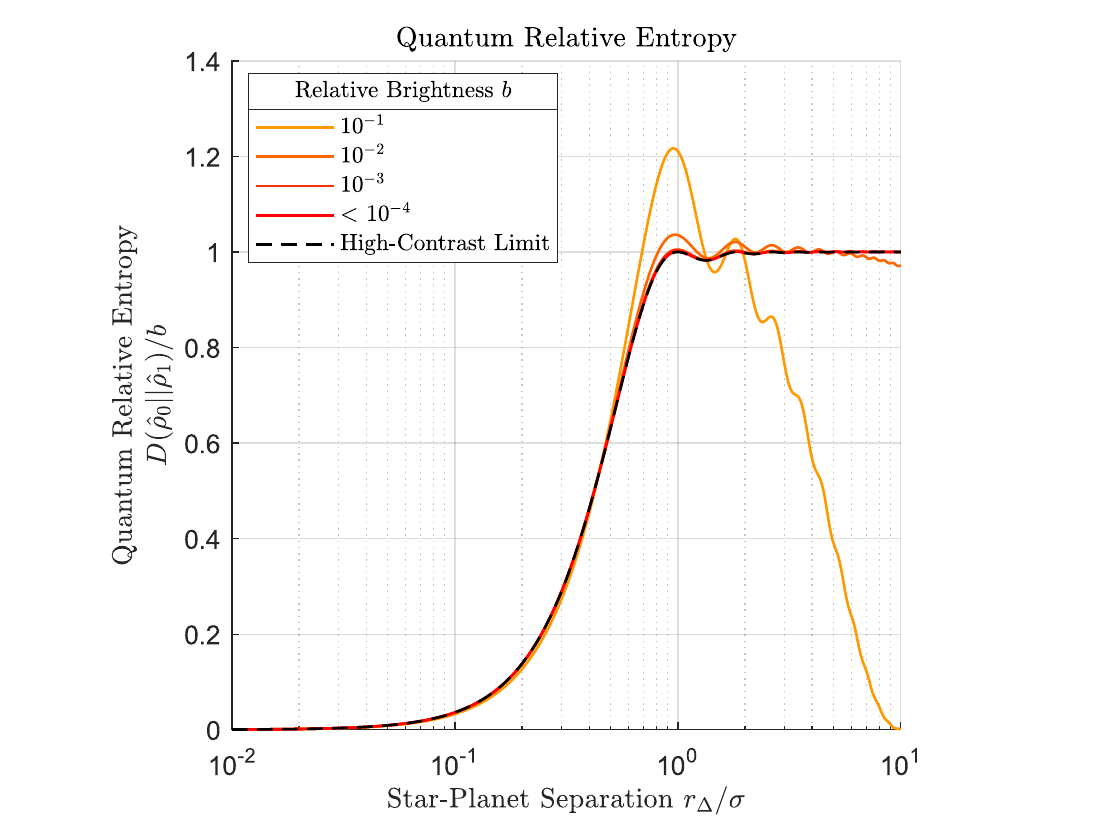}
    \caption{Plots of the quantum relative entropy under different levels of contrast. In the high-contrast limit the quantum relative entropy converges to the quantum Chernoff exponent. For the results shown, the number of photons detected is $N=10^9$ and the false-negative error tolerance is $\delta= 0.2$.}
    \label{fig: Quantum Relative Entropy}
\end{figure}

In this section we derive the quantum Stein lemma for asymmetric hypothesis testing. From Eq. \ref{eqn: Quantum Stein Lemma} we see that this involves calculating the quantum relative entropy $D(\hat{\rho}_0||\hat{\rho}_1)$, as well as the variance $V(\hat{\rho}_0||\hat{\rho}_1)$. Defining $\ket{\psi_1} = \ket{\psi_0(\mbf{r}_e)}$ and $\ket{\psi_2} =\ket{\psi_0(\mbf{r}_s)}$, one can verify that the density operator $\hat{\rho}_1$ is diagonalized by the states $\ket{\lambda_{1}},\ket{\lambda_2}$ and eigenvalues $\lambda_{1},\lambda_{2}$ given by:
\begin{align*}
    \ket{\lambda_{1}} &= a_{11}\ket{\psi_1} + a_{12}\ket{\psi_2}, & \lambda_1 = (1-\gamma)/2\\
\ket{\lambda_{2}} &=  a_{21}\ket{\psi_1} + a_{22}\ket{\psi_2} & \lambda_{2} = (1+\gamma)/2
\end{align*}
where $\gamma \equiv \sqrt{\Delta^2 + \tau^2 }$ and $\tau^2 \equiv \kappa^2(1-\Delta^2)$. The expansion coefficients are explicitly given by,
\begin{align*}
a_{11} &= N_{-}(M_{+}u_{-}-M_{-}\tau)   & a_{12} &= N_{-}(M_{+}u_{-}+M_{-}\tau)\\
a_{21} &= N_{+}(M_{+}u_{+}-M_{-}\tau)   & a_{22} &= N_{+}(M_{+}u_{+}+M_{-}\tau)
\end{align*}
with definitions
$$
u_{\pm} \equiv \Delta \pm \gamma , \qquad N_{\pm} = 1/\sqrt{\tau^2 + u_{\pm}^2},\qquad  M_\pm \equiv 1/\sqrt{2(1\pm\Delta)}
$$
Therefore the quantum relative entropy can be written as,
\begin{align*}
    D(\hat{\rho}_0||\hat{\rho}_1) &= \langle \ln\hat{\rho}_0 - \ln \hat{\rho}_1\rangle_{\hat{\rho}_0} = \Tr[\rho_0 \ln \rho_0] - \Tr[\rho_0 \ln \rho_1] = - \Tr[\rho_0 \ln \rho_1]\\
    &= (\ln \lambda_1) \Tr[\hat{\rho}_0 \dyad{\lambda_1}] - (\ln\lambda_2) \Tr[\hat{\rho}_0 \dyad{\lambda_2}]\\
    &= - \ln \lambda_1 |\braket{\psi_0}{\lambda_1}|^2 - \ln \lambda_2 |\braket{\psi_0}{\lambda_2}|^2
\end{align*}
where 
\begin{align*}
    \braket{\psi_0}{\lambda_1} &= a_{11} \Gamma_{0}(\mbf{r}_e) + a_{12}\Gamma_0(\mbf{r}_s) \\
    \braket{\psi_0}{\lambda_2} & = a_{21}\Gamma_{0}(\mbf{r}_e) + a_{22}\Gamma_{0}(\mbf{r}_s)
\end{align*}
Given the derived quantities, the variance is then straightforward to compute,
\begin{align*}
V(\hat{\rho}_0||\hat{\rho}_1) &= \langle [(\ln \hat{\rho}_0 - \ln \hat{\rho}_1) - D(\hat{\rho}_0||\hat{\rho}_1)]^2\rangle_{\hat{\rho}_0} \\
&= \langle [\ln \hat{\rho}_0 - \ln \hat{\rho}_1] ^2 \rangle_{\hat{\rho}_0 } - D(\hat{\rho}_0||\hat{\rho}_1)^2
\end{align*}
The first term evaluates to,
$$
\langle [\ln \hat{\rho}_0 - \ln \hat{\rho}_1] ^2 \rangle_{\hat{\rho}_0 } =2D(\hat{\rho}_0 || \hat{\rho}_1) + \Tr[\hat{\rho}_0(\ln \hat{\rho}_1)^2]
$$
$$
\Tr[\hat{\rho}_0(\ln \hat{\rho}_1)^2] = (\ln \lambda_1)^2 |\braket{\psi_0}{\lambda_1}|^2 + (\ln \lambda_2)^2 |\braket{\psi_0}{\lambda_2}|^2 
$$
\noindent Hence the variance is,
\begin{equation}
V(\hat{\rho}_0||\hat{\rho}_1) = 2D(\rho_0||\rho_1) + (\ln\lambda_1)^2 |\braket{\psi_0}{\lambda_1}|^2 + (\ln \lambda_2)^2|\braket{\psi_0}{\lambda_2}|^2 -D^2(\rho_0||\rho_1)
\end{equation}
In Fig. \ref{fig: Quantum Relative Entropy} we plot the quantum relative entropy for different values of exoplanet contrast and star-planet separation.


\section{Derivation of the Quantum Fisher Information Matrix}
\label{apd: QFIM Derivation}

References \cite{Prasad:2019_3DSource_APS, Prasad:2020_3DSource_IOPScience} provide the general equation of the QFIM for estimating the separation between two unbalanced point sources in 3-dimensional space. We invoke this equation for special case where both point sources are located at a resolved object plane, reducing the problem to 2-dimensional space. In the case of a real inversion-symmetric pupil function, the QFIM is given by,

\begin{equation}
\mathcal{K}_{ij} = (1-\kappa^2) \bigg[ \langle \partial_{\theta_{i}} (2\pi \mbf{u}\cdot \mbf{r}_{\Delta}) \partial_{\theta_{j}} (2\pi \mbf{u}\cdot \mbf{r}_{\Delta})  \rangle - \kappa^2 \partial_{\theta_{i}} \Delta \partial_{\theta_{j}} \Delta \bigg]
\label{eqn: Prasad QFIM}
\end{equation}
We will show that the first and second term in $\bm{\mathcal{K}}$ are associated with the single-source localization QFIM $\bm{\mathcal{K}}_{1}$ and the CFIM contribution of the fundamental mode $\bm{\mathcal{I}}_0$ respectively.  A deeper geometric interpretation of the QFIM can be found in \cite{Liu:2020} where the authors elucidate how the matrix relates to a metric on the manifold of quantum states parameterized by a general collection of variables $\bm{\theta}$.

\subsection{QFIM for Localizing a Single Point Source}
Here we derive the QFIM for localization of a single point source. Let $\mbf{r}_{\Delta}$ be the location of a single point source relative to the optical axis such that the single-photon state of the optical field is given by the pure state,
$$
\hat{\rho}_1 = \dyad{\psi_0(\mbf{r}_{\Delta})}
$$
The QFIM for pure states is given by \cite{Liu:2020},
\begin{align}
    \bm{\mathcal{K}}_{1,ij} &=  4 \Re \bigg\{ \braket{\partial_{\theta_i} \psi_{0}(\mbf{r}_{\Delta})}{\partial_{\theta_j} \psi_{0}(\mbf{r}_{\Delta})} \nonumber \\ 
    &-  \braket{\partial_{\theta_i} \psi_{0}(\mbf{r}_{\Delta})}{\psi_0(\mbf{r}_{\Delta})} \braket{\psi_0(\mbf{r}_{\Delta})}{\partial_{\theta_j} \psi_{0}(\mbf{r}_{\Delta})}\bigg\}
\label{eqn: Single-Source QFIM}
\end{align}
We will evaluate each inner product in Eq. \ref{eqn: Single-Source QFIM} separately. The first inner product evaluates to,
$$
\braket{\partial_{\theta_i} \psi_{0}(\mbf{r}_{\Delta})}{\partial_{\theta_j} \psi_{0}(\mbf{r}_{\Delta})} = \langle \partial_{\theta_i}(2\pi \mbf{u}\cdot \mbf{r}_{\Delta}) \partial_{\theta_j}(2\pi \mbf{u}\cdot \mbf{r}_{\Delta}) \rangle
$$
while the remaining inner products evaluate to zero $\braket{\partial_{\theta_i} \psi_{0}(\mbf{r}_{\Delta})}{\psi_0(\mbf{r}_{\Delta})} = 0$. Therefore, the QFIM for single-source localization is,
$$
\mathcal{K}_{1,ij} = 4 \langle \partial_{\theta_i}(2\pi \mbf{u}\cdot \mbf{r}_{\Delta}) \partial_{\theta_j}(2\pi \mbf{u}\cdot \mbf{r}_{\Delta}) \rangle
$$
which we immediately recognize as the first term in Eq. \ref{eqn: Prasad QFIM} for $\mathcal{K}_{ij}$ up to a proportionality constant. We also point out that since the derivatives in $\bm{\mathcal{K}}_1$ act on terms of first order in $\mbf{r}_{\Delta}$, the Cartesian parameterization of $\bm{\mathcal{K}}_1$ is independent of the distance $r_{\Delta}$. A dependence on $r_{\Delta}$ does arise for a polar parameterization, however it only plays the role of a geometric factor intrinsic to the Jacobian. Therefore the uncertainty patch $\sigma_{loc}$ for localizing a single point source is invariant with respect to the position of the source.

\subsection{CFIM Contribution of the Fundamental Mode}

Here again, we consider a single point source located at $\mbf{r}_{\Delta}$ relative to the optical axis. For a real-valued PSF, the probability of detecting a photon in the fundamental mode is given by,
$$
p_{\psi_0}(\mbf{r}_{\Delta}) = \Gamma^2_{\psi_0}(\mbf{r}_{\Delta}) = \Delta^2
$$
The classical Fisher information contained in the fundamental mode is therefore,
$$
\mathcal{I}_{0,ij} = \frac{\big( \partial_{\theta_i } p_{\psi_0} \big) \big(\partial_{\theta_j}p_{\psi_0} \big)}{p_{\psi_0}} = 4\big( \partial_{\theta_i } \Delta(\mbf{r}_{\Delta}) \big) \big(\partial_{\theta_j} \Delta(\mbf{r}_{\Delta}) \big)
$$
which we immediately recognize as the second term in $\mathcal{K}_{ij}$ up to a proportionality constant. Therefore, we may write the QFIM as,
$$
\bm{\mathcal{K}} = \frac{1}{4}(1-\kappa^2)\bigg(\bm{\mathcal{K}}_1 -  \kappa^2 \bm{\mathcal{I}}_0 \bigg)
$$
which is the QFIM of the star-planet separation vector $\mbf{r}_{\Delta}$ given in Eq. \ref{eqn: QFI Compact}.

\section{Post-Propagation Quantum Limits}
\label{apd: Post-Propagation Quantum Limits}

Consider a coronagraph that achieves complete nulling of the fundamental mode such that $\ket{\nu_0} = \ket{\psi_0}$ and $\eta_0 = 0$. In the high-contrast limit, it can be shown that the single-photon component of the state remaining after propagation through the coronagraph is,
$$
\hat{\rho}_1' = b\sum_{k,k'>0} \sqrt{\eta_k \eta_{k'}} \Gamma_{\nu_k}(\mbf{r}_{\Delta}) \Gamma_{\nu_{k'}}^{*}(\mbf{r}_{\Delta})\dyad{\mu_k}{\mu_{k'}} \\
 = \dyad{\Psi}
$$
where we define the \textit{non-normalized} state $$
\ket{\Psi} = \sqrt{b} \sum_{k>0} \Gamma_{\nu_k}(\mbf{r}_{\Delta}) \ket{\mu_k}
$$

The single-photon output state of the coronagraph is clearly a pure state (up to a normalization constant). Applying the definition of the QFIM for a pure state,

$$
\mathcal{K}_{ij}(\hat{\rho}_{1}') = 4\Re{ \braket{\partial_{\theta_i}\Psi}{\partial_{\theta_j}\Psi} -\braket{\partial_{\theta_i}}{\Psi}\braket{\Psi}{\partial_{\theta_j} }}
$$
and dropping the argument $\Gamma_{\nu_k}(\mbf{r}_{\Delta})\rightarrow \Gamma_{\nu_k}$ for notational convenience, we find that,
\begin{align*}
\mathcal{K}_{ij}(\hat{\rho}_{1}') &= 4\Re \bigg\{  b \sum_{k>0} \eta_{k} (\partial_{\theta_i} \Gamma_{\nu_k})^{*}(\partial_{\theta_j} \Gamma_{\nu_k}) \\
 -& b^2\sum_{k,k'>0}\eta_{k}^2 (\partial_{\theta_i}\Gamma_{\nu_k})^*(\Gamma_{\nu_k})(\Gamma_{\nu_{k'}})^{*}(\partial_{\theta_j}\Gamma_{\nu_{k'}})
    \bigg\}
\end{align*}
In the high-contrast limit, we drop the second term of order $\mathcal{O}(b^2)$ such that, 
\begin{equation}
    \mathcal{K}_{ij}(\hat{\rho}_{1}') \xrightarrow[]{b<<1} 4b \sum_{k>0} \eta_k \Re{(\partial_{\theta_i} \Gamma_{\nu_k})^{*}(\partial_{\theta_j} \Gamma_{\nu_k})} 
\end{equation}
Note that the sum is effectively the addition of classical information in the modes $\nu_{k}$ weighted by the transmission coefficients $\eta_k$. We may also express the QFIM in terms of amplitude and phase contributions:
\begin{align*}
\mathcal{K}_{ij}(\hat{\rho}_{1}')&= 4b \sum_{k>0} \eta_{k} \bigg[(\partial_{\theta_i} |\Gamma_{\nu_k}|)(\partial_{\theta_j} |\Gamma_{\nu_k}|) +  \\
&\qquad |\Gamma_{\nu_k}|^2 (\partial_{\theta_i} \arg(\Gamma_{\nu_k}))(\partial_{\theta_j} \arg(\Gamma_{\nu_k}))\bigg]
\end{align*}
Note that if the phase of the correlation coefficients $\Gamma_{\nu_k}$ are parameter independent, the second term inside the infinite summation disappears. 

\subsection{CFIM of SPADE}
Performing a SPADE measurement in the output mode basis $\mu_k$ of the coronagraph operator gives probabilities,
$$
p_{\mu_k} = \mel{\mu_{k}}{\hat{\rho}_1'}{\mu_k} = b\sum_{k>0} \eta_{k} |\Gamma_{\nu_k}|^2
$$
The CFIM of the SPADE measurement on the coronagraph channel output is,
$$
\mathcal{I}^{SPADE}_{ij}(\hat{\rho}_{1}') = 4b \sum_{k>0}\eta_{k} (\partial_{\theta_i} |\Gamma_{\nu_k}| )(\partial_{\theta_j}|\Gamma_{\nu_k}|)
$$
We observe that if the phase of $\Gamma_{\nu_k}$ has no dependence on the parameter $\mathbf{r}_{\Delta}$ then the CFIM of SPADE saturates the post-propagation QFIM of the coronagraph $\bm{\mathcal{K}}(\hat{\rho}_{1}')$.

\subsection{CFIM of Direct Imaging}
Performing a direct imaging measurement on the output state of the coronagraph channel gives single-photon detection probability,
$$
p_{\mbf{r}} = |\braket{\mbf{r}}{\Psi}| = |\Psi(\mbf{r})|^2  = |\sqrt{b} \sum_{k>0}\sqrt{\eta_k}\, \Gamma_{\nu_k} \mu_{k}(\mbf{r})|^2
$$
The CFIM of direct imaging on the coronagraph channel output is,
$$
\mathcal{I}^{DD}_{ij} = \int \frac{(\partial_{\theta_i}p_{\mbf{r}})(\partial_{\theta_j}p_{\mbf{r}})}{p_{\mbf{r}}} d\mbf{r}
$$
If $\Psi(\mbf{r})$ is either purely real or imaginary (i.e. definite parity), then we may exploit the orthogonality of the $\mu_k(\mbf{r})$'s to find that,
$$
\mathcal{I}^{DD}_{ij} = 4b \sum_{k>0} \eta_{k} (\partial_{\theta_i}\Gamma_{\nu_k})(\partial_{\theta_j}\Gamma_{\nu_k})
$$
which is equal to the post-propagation QFIM. Otherwise if $\Psi(\mbf{r})$ does not have definite parity, the direct imaging CFI is given by,
$$
\mathcal{I}^{DD}_{ij} = 4 \int (\partial_{\theta_i}|\Psi(\mbf{r})|)(\partial_{\theta_j}|\Psi(\mbf{r})|) d\mbf{r}
$$
and the post-propagation QFIM is not necessarily saturated. This last result may be derived from the relation, $\partial_{\theta} |\Psi|^2 = 2|\Psi| (\partial_{\theta} |\Psi|)$.

\section{The Fourier-Zernike Modes}
\label{apd: Fourier-Zernike Modes}

The Zernike modes $\tilde{\psi}_{nm}$ constitute a PSF-matched basis over a circular pupil,

\begin{subequations}
    \begin{align}
    \tilde{\psi}_{nm}(u,\theta) &\equiv  R_{nm}(u) \Theta_{m}(\theta) \text{circ}(u) \\
    R_{nm}(u) &\equiv \sum_{j=0}^{(n-|m|)/2} \frac{(-1)^j\sqrt{n+1}(n-j)! }{j![(n+m)/2 - j]![(n-m)/2 -j]!} u^{n-2j}\\
     \Theta_{m}(\theta) &\equiv
     \begin{cases}
     \sqrt{2} \cos(|m|\theta) & (m>0)\\
     1  &   (m = 0)\\
     \sqrt{2} \sin(|m|\theta) & (m<0)
     \end{cases}\\
     \text{circ}(u) &\equiv 
     \begin{cases}
     1, & u \leq 1\\
     0, & u > 1
     \end{cases}
    \end{align}
\label{eqn: Zernike Modes}
\end{subequations}
where the radial index range is $n = 0,1,2,\ldots,\infty$. The angular index range is $m \in \mathcal{S}_n = \{-n,-n+2,\ldots,n-2,n\}$ for a given radial index. These modes are defined to satisfy orthonormality,

$$
\int \tilde{\psi}^{*}_{nm}(u,\theta) \tilde{\psi}_{n'm'} (u,\theta) udud\theta  = \delta_{nn'}\delta_{mm'}
$$

The Fourier transform of the Zernike modes over the pupil are found in \cite{Dai:2006} to be,

\begin{equation}
\psi_{nm}(r,\phi) = i^{n+2|m|}\sqrt{n+1} \frac{J_{n+1}(2\pi r)}{\sqrt{\pi} r}\Theta_{m}(\phi)
\label{eqn: Fourier-Zernike Modes}
\end{equation}

\section{SPADE CFIM in the Fourier-Zernike Modes}
\label{apd: FZ CFIM Calculations}

In this section we derive the CFIM for a SPADE system configured to sort the Fourier-Zernike modes. In the first section, we express the complete CFIM free of approximations. In the second section, we consider the CFIM in the high contrast limit and prove that it equals the high-contrast QFIM.

\begin{figure}
    \centering
    \includegraphics[width=\linewidth]{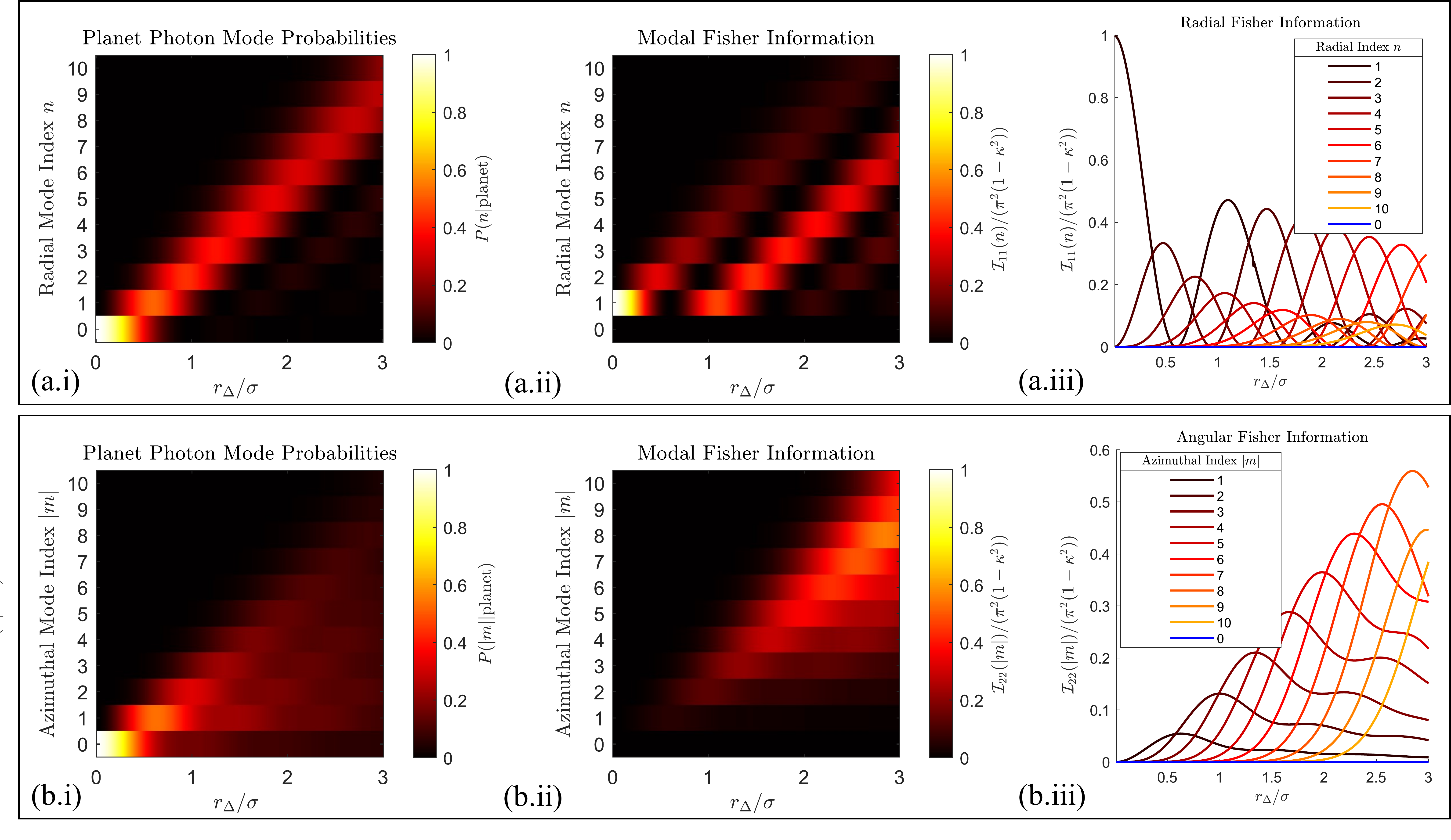}
    \caption{Decomposition of the Classical Fisher Information by Zernike mode over different star-planet separations with relative brightness $b = 10^{-9}$ and orientation angle $\phi_{\Delta} = 45^{o}$. Sub-panels \textbf{(a.i)} and \textbf{(b.i)} show the evolution of the modal coupling probabilities for photons emitted by the exoplanet as a function of the star-planet separation. We see that the coupling probabilities spread to higher-order modes as the star-planet separation increases. Sub-panels \textbf{(a.ii-iii)} and \textbf{(b.ii-iii)} depict how the concentration of information about the radial parameter $r_{\Delta}$ and angular parameter $\phi_{\Delta}$ evolves with respect to the star-planet separation over Zernike modes with radial index $n$ and angular index $|m|$, respectively. The fundamental mode (index pair $n=0,m=0$) carries virtually no information about the separation parameters.}
    \label{fig: FZ Mode CFIs}
\end{figure}

\subsection{Complete SPADE CFIM}

 The polar coordinates for the star and planet vectors are given by,

\begin{align*}
(r_s,\phi_s) &= \big( b r_{\Delta}, \phi_{\Delta} + \pi \big) \\
(r_e,\phi_e) &= \big( (1-b)r_{\Delta},\phi_{\Delta} \big)
\end{align*}
where $r_{\Delta} \in [0,\infty)$ and $\phi_{\Delta} \in [0,2\pi)$. Let $P_{nm} = (1-b)p_{nm}(r_s,\phi_s) + b p_{nm}(r_e,\phi_e)$ be the probability of detecting a photon in Fourier-Zernike mode $n,m$ under the star-planet configuration where
$$
p_{nm} (r,\phi)=  |\Gamma_{nm}(r,\phi)|^2 = \frac{1}{\pi}\bigg[ \sqrt{n+1} \frac{J_{n+1}(2\pi r)}{\sqrt{\pi}r} \Theta_{m}(\phi)\bigg]^2
$$
is the probability of detecting a photon in mode $n,m$ if a single point source is located at position $\mbf{r}$. The CFIM contribution of each mode is given by,
$$
\mathcal{I}^{r\phi}_{ij}[n,m] =  \frac{\big( \partial_{\theta_i} P_{nm} \big) \big(\partial_{\theta_j} P_{nm} \big) }{P_{nm}}
$$
where $\theta_i \in \{r_{\Delta},\phi_{\Delta}\}$. Thus, we begin by calculating the partial derivatives of the single-source mode probabilities,

\begin{subequations}
\begin{align}
\label{eqn: Derivative FZ Radial}
    p^{(r)}_{nm}(r,\phi) &\equiv \partial_{r} p_{nm}(r,\phi) =  2 \bigg(\frac{J_{n+1}(2 \pi r)}{r}\bigg) \nonumber \\
    \times & \bigg[ J_{n-1}(2 \pi r)- J_{n-3}(2 \pi r)\bigg] \bigg( \Theta_{m}(\phi)\bigg)^2 \\
\label{eqn: Derivative FZ Angular}
    p^{(\phi)}_{nm}(r,\phi) &\equiv \partial_{\phi} p_{nm}(r,\phi) = 4 \bigg( \frac{J_{n+1}(2\pi r)}{ r} \bigg)^2  \nonumber \\
    \times &  (n+1) (-m) \cos(|m|\phi) \sin(|m|\phi)
\end{align}
\end{subequations}

Applying the chain rule, the partial derivatives of the total mode probability is,

\begin{align*}
\partial_{r_{\Delta}} P_{nm} &= (1-b) \partial_{r_{\Delta}} p_{nm}(b r_{\Delta}, \phi_{\Delta} + \pi) + b \partial_{r_{\Delta}} p_{nm}((1-b) r_{\Delta}, \phi_{\Delta}) \\
&=b(1-b)\bigg[p^{(r)}_{nm}(br_{\Delta},\phi_{\Delta}+\pi) + p^{(r)}_{nm}((1-b)r_{\Delta},\phi_{\Delta}) \bigg]\\
\partial_{\phi_{\Delta}} P_{nm} &= (1-b) \partial_{\phi_{\Delta}} p_{nm}(b r_{\Delta}, \phi_{\Delta} + \pi) + b \partial_{\phi_{\Delta}} p_{nm}((1-b) r_{\Delta}, \phi_{\Delta}) \\
 &= (1-b) p^{(\phi)}_{nm}(b r_{\Delta}, \phi_{\Delta} + \pi) + b p^{(\phi)}_{nm}((1-b) r_{\Delta}, \phi_{\Delta}) \\
\end{align*}

We refrain from substituting all elements into one equation as the resulting expression would be excessively long. However, the equations provided enable quick computation of the Fisher information contribution of each Zernike mode $\mathcal{I}_{ij}^{(r\phi)}[n,m]$. Drawing attention to Eq. \ref{eqn: Derivative FZ Angular}, we see that $p^{(\phi)}_{nm}(r,\phi) = 0$ for all $n,m$  when $\phi$ is an integer multiple of $\frac{\pi}{2}$. This implies that the CFI of the angular coordinate is zero when the separation axis of the star-planet system exactly coincides with the x-axis or the y-axis. These apparent 'singularities' in the CFI can be neglected in practice because the probability of exact alignment along the x and y axes is infinitesimally small. That said, the disappearing CFI has non-negligible effects when considering a finite Zernike mode basis that is truncated at some arbitrarily large order. We find that a good rule of thumb for simulating the localization of point sources near the x and y axes is to use a truncated Zernike mode basis with the max radial being bounded by $n \gtrsim \lceil \frac{\pi}{2 |\phi_{min}|} \rceil$ where $\phi_{min}$ is the minimum orientation angle (in radians) that the star-planet separation axis makes with either x or y axes. Alternatively, one may choose a lower mode truncation order and simply rotate the basis by $\pi/4$ radians halfway through the photon collection period to cover the singularities near the x,y axes. 

\subsection{SPADE CFIM in the High-Contrast Limit}

In the high contrast limit where $b\rightarrow 0, \mbf{r}_s \rightarrow 0,$ and $\mbf{r}_{e}\rightarrow \mbf{r}_{\Delta}$, the mode probabilities approach

$$
P_{k} \rightarrow (1-b)p_{k}(0) + b p(\mbf{r}_{\Delta})
$$

where we have introduced the OSA/ANSI standard linear index of the Zernikes $k = (n(n+2)+m)/2$ for later convenience. Critically, for a PSF-matched SPADE basis, we have $p_{k}(0) = \delta_{k0}$. Moreover, inspecting Eq. \ref{eqn: Fourier-Zernike Modes}, we see that the Fourier-Zernike modes are either purely imaginary or purely real based on the parity of $n+2|m|$. Since this is independent of the argument $\mbf{r}$, we may re-define $\psi_k(\mbf{r}) \leftarrow i^{-(n+2|m|)} \psi_k(\mbf{r})$ without loss of generality to make the Fourier-Zernike modes purely real. This allows us to write the magnitude squared of the correlation functions $|\Gamma_k(\mbf{r})|^2 = \Gamma_k(\mbf{r})^2$ such that the CFIM terms adopt the convenient form,

$$
\mathcal{I}_{k,ij} \rightarrow 4b^2 \Gamma^2_{k}(\mbf{r}_{\Delta}) \, \frac{\bigg( \partial_{\theta_i} \Gamma_k(\mbf{r}_{\Delta})\bigg)\bigg( \partial_{\theta_j} \Gamma_k(\mbf{r_{\Delta}})\bigg)}{\delta_{k0} +b \Gamma^2_k(\mbf{r}_{\Delta})} 
$$

Note that the presence of the Kronecker Delta in the denominator means the information contribution of the fundamental mode $k=0$ scales as $b^2$. Meanwhile, the information contribution of higher-order modes $k>0$ scales as $b$. Thus, in the high-contrast limit the SPADE CFIM approaches

\begin{equation}
\mathcal{I}_{ij} \rightarrow 4b \sum_{k>0} \bigg( \partial_{\theta_i}\Gamma_{k}(\mbf{r}_{\Delta})\bigg)\bigg(\partial_{\theta_j}\Gamma_{k}(\mbf{r}_{\Delta})\bigg)
\label{eqn: High-Contrast FZ CFIM}
\end{equation}

\subsection{Asymptotic Optimality of SPADE}

 To prove the optimality of SPADE for localization, we show that Eq. \ref{eqn: High-Contrast FZ CFIM} is equal to the high-contrast QFIM of Eq. \ref{eqn: QFIM Polar HC Circular Aperture} for a circular aperture. We note that this derivation also constitutes a proof for the quantum optimality of the Zernike modes in the canonical balanced two-source localization problem involving a circular aperture. We begin by evaluating the diagonal terms of the CFIM. Consider the infinite sum of radial derivatives

$$
\sum_{k=0}^{\infty}  (\partial_{r} \Gamma_{k}(r, \phi))^2
$$

The radial derivate of the Fourier-Zernike modes are,

$$
\partial_{r} \Gamma_{nm}(r, \phi) = \frac{\pi}{\sqrt{n+1}}\bigg[J_{n-1}(2\pi r) - J_{n+3}(2\pi r) \bigg]\Theta_{m}(\phi) 
$$
such that the infinite sum resolves,
$$
\sum_{n=0}^{\infty} \sum_{m\in \mathcal{S}_n}  (\partial_{r} \Gamma_{nm}(r, \phi))^2 = \pi^2 
$$
which can be shown by invoking the identity, $\sum_{n=0}^{\infty} \bigg[J_{n-1}(x) - J_{n+3}(x)\bigg]^2 = 1 \,, \forall x\geq0 $. Next consider the infinite sum of angular derivatives
$$
\sum_{k=0}^{\infty}  (\partial_{\phi} \Gamma_{k}(r, \phi))^2
$$
The angular derivative of the Fourier-Zernike modes are
$$
\partial_{\phi} \Gamma_{nm}(r, \phi) = \sqrt{n+1}\frac{J_{n+1}(2\pi r)}{\pi r} \bigg[ -m\Theta_{-m}(\phi)\bigg]
$$
such that the infinite sum resolves,
$$
\sum_{n=0}^{\infty} \sum_{m\in \mathcal{S}_n}  (\partial_{\phi} \Gamma_{nm}(r, \phi))^2 = (\pi r)^2
$$
which can be shown by invoking the identity $\frac{4}{3} \sum_{n=0}^{\infty} n(n+2) \bigg[J_{n}(x) + J_{n+2}(x)\bigg]^2  = x^2\,, \forall x\geq0 $. Applying these terms to the high-contrast CFI of Eq. \ref{eqn: High-Contrast FZ CFIM}, we see that 

\begin{align}
    \mathcal{I}^{(r\phi)}_{11} &= 4b\bigg(\pi^2  - \big(\partial_{r_{\Delta}} \Gamma_0(r_{\Delta},\phi_{\Delta})\big)^2 \bigg) \nonumber \\
    &= 4 \pi^2 b \bigg( 1 - \bigg( \frac{2 J_{2}(2\pi r_{\Delta})}{\pi r_{\Delta}}\bigg)^2 \bigg) \\
    \mathcal{I}^{(r\phi)}_{22} &= 4b\bigg((\pi r_{\Delta})^2 - \big(\partial_{\phi_{\Delta}} \Gamma_0(r_{\Delta},\phi_{\Delta})\big)^2 \bigg) \nonumber \\ 
    &= 4 \pi^2 b r_{\Delta}^2
\end{align}
which are precisely the diagonal terms of the high-contrast QFIM shown in Eq. \ref{eqn: QFIM Polar HC Circular Aperture}. The off-diagonal terms of the CFIM are equal to zero,
$$
\mathcal{I}^{(r \phi)}_{12} = \mathcal{I}^{(r \phi)}_{21} = 4b  \sum_{n=0}^{\infty}\sum_{m \in \mathcal{S}_n} \partial_{r}\Gamma_{nm} \partial_{\phi}\Gamma_{nm} = 0
$$
which can be easily shown by recognizing that the sums over the angular indices evaluate to zero. This concludes the proof.

\section{Asymptotic Optimality of the Perfect Coronagraph}
\label{apd: Quantum-Optimal PC Proof}

In this section we will prove that the Perfect Coronagraph (PC) asymptotically approaches the QFIM in the high-contrast regime $b<<1$ between the star and the exoplanet. The coronagraph operator for the PC can be written as $\hat{C}_1 = \hat{I}_1-\dyad{\psi_0}$. The field state post-nulling of the fundamental PSF mode is given by,
\begin{align*}
\hat{\rho}' &= (\hat{I}_1-\dyad{\psi_0}) \hat{\rho} (\hat{I}_1-\dyad{\psi_0})\\
&= \hat{\rho} + \mel{\psi_0}{\hat{\rho}}{\psi_0}\dyad{\psi_0} - \dyad{\psi_0}\hat{\rho} -  \hat{\rho}\dyad{\psi_0} 
\end{align*}

where $\ket{\psi_0}$ is the single photon state in the fundamental mode. Our goal is to determine the CFI for a direct-imaging measurement $\{ \hat{\Pi}_{\mathbf{r}} = \dyad{\mathbf{r}} :\mbf{r}\in \mathbb{R}^2\}$ on the nulled density operator. First we find the probability of detecting a photon at location $\mathbf{r}$ on the image plane:
\begin{align*}
p(\mathbf{r}) = &\mel{\mbf{r}}{\hat{\rho}'}{\mbf{r}} \\
= &\rho(\mbf{r}) + \mel{\psi_0}{\hat{\rho}}{\psi_0}|\psi_{0}(\mbf{r})|^2 \\
&-  \psi_0(\mbf{r}) \mel{\psi_0}{\hat{\rho}}{\mbf{r}} - \psi^{*}_0(\mbf{r}) \mel{\mbf{r}}{\hat{\rho}}{\psi_0}\\
= &\rho(\mbf{r}) + \mel{\psi_0}{\hat{\rho}}{\psi_0}|\psi_{0}(\mbf{r})|^2 - 2\Re{ \psi_0(\mbf{r}) \mel{\psi_0}{\hat{\rho}}{\mbf{r}} }
\end{align*}
which expands to
 
\begin{equation*}
\begin{split}
p(\mathbf{r}) = &(1-b)|\psi_0(\mbf{r}-\mbf{r}_s)|^2 + b |\psi_0(\mbf{r}-\mbf{r}_e)|^2 \\
& + |\psi_0(\mbf{r})|^2 \bigg((1-b) |\Gamma_0(\mbf{r}_{s})|^2 +  b |\Gamma_0(\mbf{r}_{e})|^2 \bigg) \\
& -2 \Re \bigg\{\psi_0(\mbf{r}) \bigg( (1-b)\Gamma_0(\mbf{r}_s) \psi^{*}_0(\mbf{r}-\mbf{r}_s) \\
&+ b \Gamma_0(\mbf{r}_e) \psi^{*}_0(\mbf{r}-\mbf{r}_e) \bigg) \bigg\}
\end{split}
\end{equation*}

In the limit of high contrast $b<<1$ where $\mbf{r}_s \rightarrow 0$ and $\mbf{r}_{e}\rightarrow \mbf{r}_{\Delta}$, the probability is well-approximated by,

\begin{equation*}
\begin{split}
p(\mathbf{r}) \approx & 2(1-b) |\psi_0(\mbf{r})|^2 + b |\psi_0(\mbf{r}-\mbf{r}_{\Delta})|^2 + b |\psi_0(\mbf{r})|^2 
 |\Gamma_0(\mbf{r}_{\Delta})|^2\\
& -2 \Re{\psi_0(\mbf{r}) \bigg( (1-b) \psi^{*}_0(\mbf{r}) + b \Gamma_0(\mbf{r}_{\Delta}) \psi^{*}_0(\mbf{r}-\mbf{r}_{\Delta}) \bigg)}
\end{split}
\end{equation*}
Assuming a real-valued PSF, this further reduces to

\begin{align*}
p(\mathbf{r}) \approx & b \bigg[\psi^{2}_0(\mbf{r} - \mbf{r}_{\Delta}) +  \psi^2_0(\mbf{r})
\Gamma^2_0(\mbf{r}_{\Delta}) - 2 \psi_0(\mbf{r})\Gamma_0(\mbf{r}_{\Delta}) \psi_0(\mbf{r}-\mbf{r}_{\Delta}) \bigg]\\
\approx & b \bigg[\psi_0(\mbf{r} - \mbf{r}_{\Delta}) - \Gamma_0(\mbf{r}_{\Delta})\psi_0(\mbf{r}) \bigg]^2
\end{align*}
Let us now define the real-valued proxy correlation function 

$$
\Gamma(\mbf{r};\mbf{r}_{\Delta}) \equiv \sqrt{b} \bigg[\psi_0(\mbf{r} - \mbf{r}_{\Delta}) - \Gamma_0(\mbf{r}_{\Delta}) \psi_0(\mbf{r}) \bigg]
$$
such that $p(\mbf{r}) \approx \Gamma^2(\mbf{r};\mbf{r}_{\Delta})$. 
Doing so allows us to write the CFI matrix in the convenient form,
$$
\mathcal{I}_{ij} \approx \int d^2r \frac{(\partial_{\theta_i}  \Gamma^2)(\partial_{\theta_j}  \Gamma^2)}{\Gamma^2} = 4 \int d^2r (\partial_{\theta_i}  \Gamma) (\partial_{\theta_j}  \Gamma)
$$

Note that $\Gamma(\mbf{r};\mbf{r}_{\Delta})$ is proportional to the optical field induced by a shifted PSF minus its component in the fundamental mode. Moreover, the shifted PSF admits an expansion in a PSF-matched orthonormal basis,

$$
\psi_0(\mbf{r}-\mbf{r}_{\Delta}) = \sum_{k=0}^{\infty}\Gamma_k(\mbf{r}_{\Delta}) \psi_{k}(\mbf{r})
$$
such that we may express the proxy correlation function as,
$$
\Gamma(\mbf{r};\mbf{r}_{\Delta}) = \sqrt{b} \sum_{k>0} \Gamma_k(\mbf{r}_{\Delta}) \psi_{k}(\mbf{r}) 
$$
Substituting this expression into the CFI and invoking the orthogonality of the modes, we have

$$
\mathcal{I}_{ij} \rightarrow 4b \sum_{k>0} \big( \partial_{\theta_i} \Gamma_{k}(\mbf{r}_{\Delta}) \big) \big( \partial_{\theta_j} \Gamma_{k}(\mbf{r}_{\Delta}) \big)
$$
which we immediately recognize as Eq. \ref{eqn: High-Contrast FZ CFIM}, the CFIM of SPADE in the high-contrast limit. We proved  Eq. \ref{eqn: High-Contrast FZ CFIM} approaches QFIM in Appendix \ref{apd: FZ CFIM Calculations}.2. Hence the PC is quantum optimal.

\end{document}